\documentclass[a4paper,fleqn,usenatbib]{mnras}

\usepackage{newtxtext,newtxmath}
\usepackage[T1]{fontenc}
\usepackage{ae,aecompl}

\usepackage{graphicx}	% Including figure files
\usepackage{amsmath}	% Advanced maths commands
\numberwithin{equation}{section} % eq number 3.1, 3.2 etc
\usepackage{amssymb}	% Extra maths symbols
\usepackage[separate-uncertainty = true,multi-part-units=brackets]{siunitx}      % SI units
\usepackage{wasysym}      % solar symbol
\usepackage{commath}      % format d in differentials
\usepackage{physics}  
\usepackage[noabbrev]{cleveref}     % clever replacement for ref (eq's, sec's, etc)

\newcommand*\mean[1]{\langle #1 \rangle}
\newcommand{\torus}{\textsc{torus}} % small caps for torus
\newcommand{\HII}{\ion{H}{II}} % H II formatting
\newcommand{\subnu}{_{\nu}}

\crefname{figure}{Fig.}{Figs.}
\Crefname{figure}{Fig.}{Figs.}
\crefname{table}{Table}{Tables}
\newcommand{\solar}{\ensuremath{_{\astrosun}}} 
\DeclareSIUnit\erg{erg}
\DeclareSIUnit\dyne{dyne}
\DeclareSIUnit\msol{M\solar{}}
\DeclareSIUnit\lsol{L\solar{}}
\DeclareSIUnit\rsol{R\solar{}}
\DeclareSIUnit\yr{yr}
\DeclareSIUnit\pc{pc}
\DeclareSIUnit\str{str}
\DeclareSIUnit\jy{Jy}
\DeclareSIUnit\micron{\micro\metre}
\title[Massive star feedback with Monte Carlo RHD]{Modelling massive-star feedback with Monte Carlo radiation hydrodynamics: photoionization and radiation pressure in a turbulent cloud}

\author[A. Ali, T. J. Harries, T. A. Douglas]{
Ahmad Ali,\thanks{E-mail: aali@astro.ex.ac.uk}
Tim J. Harries,
Thomas A. Douglas
\\
Department of Physics and Astronomy, University of Exeter, Stocker Road, Exeter EX4 4QL, United Kingdom
}

\date{Accepted XXX. Received YYY; in original form ZZZ}

\pubyear{2018}

\begin{document}
%\begin{NoHyper} 
\label{firstpage}
\pagerange{\pageref{firstpage}--\pageref{lastpage}}
\maketitle

\begin{abstract}
%This is a simple template for authors to write new MNRAS papers.
%The abstract should briefly describe the aims, methods, and main results of the paper.
%It should be a single paragraph not more than 250 words (200 words for Letters).
%No references should appear in the abstract. 
We simulate a self-gravitating, turbulent cloud of \SI{1000}{\msol} with photoionization and radiation pressure feedback from a \SI{34}{\msol} star. We use a detailed Monte Carlo radiative transfer scheme alongside the hydrodynamics to compute photoionization and thermal equilibrium with dust grains and multiple atomic species. Using these gas temperatures, dust temperatures, and ionization fractions, we produce self-consistent synthetic observations of line and continuum emission. 
We find that all material is dispersed from the $(\SI{15.5}{\pc})^3$ grid within \SI{1.6}{\mega\yr} or 0.74 free-fall times. Mass exits with a peak flux of \SI{2e-3}{\msol\per\yr}, showing efficient gas dispersal. The model without radiation pressure has a slight delay in the breakthrough of ionization, but overall its effects are negligible. 85 per cent of the volume, and 40 per cent of the mass, become ionized -- dense filaments resist ionization and are swept up into spherical cores with pillars that point radially away from the ionizing star.
We use free-free emission at \SI{20}{\cm} to estimate the production rate of ionizing photons. This is almost always underestimated: by a factor of a few at early stages, then by orders of magnitude as mass leaves the volume. We also test the ratio of dust continuum surface brightnesses at 450 and \SI{850}{\micron} to probe dust temperatures. This underestimates the actual temperature by more than a factor of 2 in areas of low column density or high line-of-sight temperature dispersion; the \ion{H}{II} region cavity is particularly prone to this discrepancy. However, the probe is accurate in dense locations such as filaments.

\end{abstract}


%%%%%%%%%%%%%%%%%%%%%%%%%%%%%%%%%%%%%%%%%%%%%%%%%%

%%%%%%%%%%%%%%%%% BODY OF PAPER %%%%%%%%%%%%%%%%%%

\section{Introduction}

Stars are formed in clusters within giant molecular clouds (GMCs), with observations in the Galaxy showing that the star formation efficiency (SFE), the fraction of the total mass in stars as opposed to gas, is a few per cent \citep{lada2003}. Numerical simulations of star formation tend to overestimate this proportion unless they invoke feedback mechanisms to drive down the SFE \citep{krumholz2015a}, for example by introducing thermal feedback which can prevent fragmentation \citep{krumholz2007,bate2009} or by removing reservoirs of gas that might otherwise condense into stars. This then effects the state of the stellar cluster once gas has been fully dispersed, as the exposed cluster may expand, lose stars, or become entirely unbound, depending on the SFE and dispersal timescale \citep{lada1984}. 

Gas dispersal is thought to be driven by massive stars (of spectral type O or B and mass $>\SI{8}{\msol}$), since they emit ionizing radiation which heats gas to \SI{e4}{\K}, increasing thermal pressure and driving expansion on length scales large enough to disrupt GMCs. The effect of ionizing feedback can be positive or negative with regards to the SFE. \citet{elmegreen1977} and \citet{whitworth1994} showed using analytical arguments that the material collected together by shocks from expanding ionized gas can drive material into dense layers which then gravitationally fragments to form new stars. This is supported by numerical models of clouds irradiated by ionizing stars internally \citep{dale2007} as well as externally, with \citet{dale2007a} finding that feedback caused some stars to form earlier compared to control runs without feedback; furthermore, it caused the formation of new stars that would not have formed without feedback. On the other hand, the same simulation also displayed negative effects, with star formation being delayed for some objects, and overall the increase in SFE was small. Similarly, simulations by \citet{walch2013} found that although triggering was effective on small timescales, on large timescales the SFE was reduced due to the dispersal of gas. This was also supported by \citet{geen2017}, whose models showed low SFEs consistent with observations of nearby clouds in the Galaxy. 

Photoionization is not the only feedback process in play, however. As photons interact with gas and dust grains they exert a radiation pressure which can change the morphology of \HII{} regions and sculpt gas into shells \citep{draine2011}, or reduce the SFE by clearing cavities \citep{agertz2013}. Massive stars also launch winds at high velocities ($>\SI{1000}{\kilo\m\per\s}$), shocking gas to high temperatures ($>\SI{e7}{K}$; \citealp{krumholz2015a}), and this may propagate out into the surrounding ISM \citep{lopez2011}. At the end of their lifetime, after a few Myr, massive stars explode as supernovae (SNe), injecting energy and momentum into the surroundings cleared out by feedback during the main-sequence  \citep{yorke1989,rogers2013}. The relative importance of these feedback processes is still not certain. In observational studies of the \HII{} region 30 Doradus, \citet{lopez2011} concluded that direct radiation pressure (from stellar photons) dominated over thermal pressure from ionized gas and wind-shocked gas, as well as indirect (dust-processed) radiation pressure. In 32 other \HII{} regions, \citet{lopez2014} observed that the ionized gas pressure was dominant, with two regions having a similar level of indirect radiation pressure, and all had significantly lower direct radiation pressure. 

Analytical and numerical models provide a way to constrain the impact of feedback and much work has been done towards this at different length scales and time scales. \citet{whitworth1979} and \citet{tenorio-tagle1979} modelled the dispersal of ionized gas via champagne flows in 1D, and this was built upon by \citet{yorke1989} using 2D simulations with supernovae exploding into the \HII{}-region cavity. Models by \citet{matzner2002} concluded that expanding \HII{} regions had a greater impact than stellar winds and supernovae in driving turbulence within GMCs. \citet{krumholz2009a} determined that the impact of radiation pressure in the dynamics of \HII{} regions increased with the number of massive stars and luminosity, and this requires high-surface density clouds \citep{fall2010}. \citet{peters2010} studied the growth of compact \HII{} regions as protostars developed in a cluster, while \citet{dale2011} simulated larger-scale GMCs, first with ionizing feedback, then later combining with stellar winds; the latter had less of an effect on the dynamics of large clouds, but in small clouds sculpted cavities through which ionizing radiation could leak \citep{dale2014}. \citet{rogers2013} simulated stellar winds and SNe and found that dense gas was largely resistant, with energy blowing out through low-density channels. More recently, \citet{geen2015} and \citet{geen2016} combined photoionization with SNe, with the latter's deposition of momentum depending on the number of ionizing sources pre-SN and the extent to which they had dispersed gas.

%%%%% SYNTHETIC OBSERVATIONS

While numerical simulations such as these provide crucial information on the dynamics of star-forming regions, they must still be analysed in the same way that observers view real clouds to properly compare simulation with observation. It is therefore necessary to produce synthetic observations from the hydrodynamical models. 
This is a growing field, with examples such as \citet{kurosawa2004}, who used the Monte Carlo radiative transfer (MCRT) code \torus{} to produce synthetic spectral energy distributions (SEDs) and Spitzer far-infrared observations of a completed Smoothed Particle Hydrodynamics (SPH) simulation of accretion disks in a low-mass star forming region. They used this data to verify the robustness of disk identification diagnostics.
\citet{haworth2012a} synthesised metal forbidden-line images to calculate electron densities and temperatures of an \HII{} region. 
Similarly, \citet{koepferl2017} used \textsc{hyperion} \citep{robitaille2011} to extract observations of dust continuum from SPH models of massive star feedback by \citet{dale2014}. This allowed the testing of diagnostics used to calculate densities, temperatures and star formation rates \citep{koepferl2017a,koepferl2017b}. Models by \citet{dale2012} were also post-processed by \citet{hubber2016} using \textsc{mocassin}, another MCRT photoionization code, extracting emission lines from H, He, and metals. 

The standard in modelling star forming regions has so far been to carry out a radiation hydrodynamics (RHD) calculation with a simplified radiative transfer (RT) scheme, for example ray-tracing to find the ionized Str{\"o}mgren volume and setting the temperature inside to \SI{e4}{K} \citep[e.g.][]{walch2012,dale2012}. Snapshots from these models are then post-processed with a more detailed RT scheme to synthesise observations. However, since radiation and dynamics are physically intertwined every step of the way -- temperature changes pressure which causes motion which sets a new condition for radiation, and so on -- it would be more accurate to use the detailed microphysical prescription at every timestep of the RHD calculation. The resulting parameters, such as ionization states and temperatures, can then be fed into the synthetic observation processing without having to make post-hoc assumptions. This is what we set out to do in this paper. We describe our numerical methods in \cref{sec:numericalmethods} and set out the initial conditions in 
\cref{sec:initialconditions}. We present the results of the RHD model in \cref{sec:rhdresults}, discuss them in \cref{sec:discussion}, and show synthetic observations in \cref{sec:syntheticobservations}. Finally we summarise and conclude in \cref{sec:conclusions}.

\section{Numerical methods}
\label{sec:numericalmethods}
We use the Monte Carlo (MC) radiative transfer (RT) and hydrodynamics (HD) code \torus{} as described by \citet{harries2015}. Similar MCRT methods are used by \textsc{mocassin} \citep{ercolano2003} and \citet{wood2004}, both of whom include detailed photoionization balance and dust physics. In \torus{}, the MCRT is coupled with a hydrodynamics module as described in \cref{sec:hydro}, meaning an RHD simulation is carried out self-consistently. 

The MCRT algorithm is based on \citet{lucy1999}. At the beginning of a timestep of duration $\Delta t$, the total stellar luminosity $L$ is split into a total of $N$ packets, each representing a bundle of photons of a particular frequency. Each packet has an energy
\begin{equation}
	\label{eq:epsilon}
	\epsilon_i = w_i \frac{L \Delta t}{N}
\end{equation}
with packets weighted with a factor $w_i$ (whose sum is normalised to 1) depending on the frequency of the photons, such that a packet containing photons of a high frequency effectively has fewer photons and vice versa. Packets propagate through the grid with randomly sampled path lengths $\ell$ between events representing absorption, scattering, or cell-boundary crossings. After absorption events, packets are re-emitted with a frequency sampled from a probability density function constructed from the cell emissivities, and are appropriately re-weighted with a new $w_i$; thus \torus{} uses a polychromatic treatment of both the stellar and diffuse radiation fields. If each path takes a time $\delta t = \ell / c$, then for a total timestep $\Delta t$ the path contributes an energy $\epsilon_i \delta t / \Delta t = \epsilon_i \ell / c \Delta t$ to the radiation field. The total energy density is then
\begin{equation}
	\label{eq:energydensitymc}
	\dif u\subnu{} = \frac{1}{c \Delta t V} \sum \epsilon \ell 
\end{equation}
where the sum is over events, and this is proportional to the mean intensity $J\subnu{}$ via 
\begin{equation}
	\label{eq:energydensity}
	\dif u\subnu{} = \frac{4 \pi J_{\nu}}{c} \dif{}\nu
\end{equation}
This allows the path length algorithm to be used for applications of radiative equilibrium and photoionization equilibrium.

\subsection{Photoionization balance}
\label{sec:photoionization} 

We use the photoionization algorithm described by \citet{haworth2012}, which is similar to \citet{ercolano2003}. Equating the rates of photoionization and recombination yields the ratio of number densities $n$ of successive ionization states $i$ of species $X$ \citep{osterbrock06}:
\begin{equation}
	\label{eq:ionbalance}
	\frac{n(X^{i+1})}{n(X^i)} = \frac{1}{n_e \alpha(X^i,T)} \int_{\nu_I}^\infty \frac{a\subnu{}(X^i) 4 \pi J\subnu{} }{h\nu}  \dif \nu 
\end{equation}
where $n_e$ is the electron number density, $\alpha$ is the total recombination coefficient to all energy levels, $a\subnu{}$ is the absorption cross-section, and $\nu_I$ is the ionization frequency for species $X^{i}$. Making use of \cref{eq:energydensitymc,eq:energydensity}, the mean intensity $J\subnu$ can be replaced with Monte Carlo estimators to give
\begin{equation}
	\label{eq:ionbalancemc}
	\frac{n(X^{i+1})}{n(X^i)} = \frac{1}{n_e \alpha(X^i)} \frac{1}{V\Delta t} \sum_{\nu_I}^\infty \frac{a\subnu{}(X^i) \epsilon \ell}{h\nu} 
\end{equation}

\citet{haworth2015} investigated the relative importance of different microphysics on the expansion of \HII{} regions. As Lyman continuum photons are absorbed by dust grains and re-emitted in the infrared, \citeauthor{haworth2015} concluded that dust is important in the early stages of the evolution, as the number $N_\textnormal{Ly}$ that goes into ionizing the gas is reduced; the effect is to reduce the Str{\"o}mgren radius $r_\textnormal{S}$ by about 10 per cent, since $r_\textnormal{S} \propto N_\textnormal{Ly}^{1/3}$. \citeauthor{haworth2015} also tested the sensitivity to temperature of the \HII{} region expansion. By treating the radiation field polychromatically, including the contribution from the diffuse gas as well as stars, and also including cooling from helium and metals, temperatures are reduced such that the ionized radius at later times is reduced by about 10 per cent compared to models which only consider hydrogen and a monochromatic, on-the-spot treatment. Therefore we use the full treatment in these cluster simulations.

\subsection{Thermal balance}
\label{sec:thermalbalance} 
Thermal balance is calculated in the same way as \citet{haworth2015}. We calculate separate temperatures for the gas and dust in our models, which are only coupled by a heat-exchange rate to account for collisions.

Dust heating is calculated assuming radiative equilibrium using the \citet{lucy1999} algorithm. The rates at which gas absorbs and emits radiative energy are, respectively,
\begin{align}
    \label{eq:Adot}
	\dot{A} &= 4 \pi \int_0^{\infty} k_{\nu} J_{\nu} \dif{}\nu  \\
	\label{eq:Edot}
	\dot{E} &= 4\pi \int_0^{\infty} k_{\nu} B_{\nu}(T_d) \dif{}\nu
\end{align}
where $k\subnu$ is the opacity per unit length. \Cref{eq:Adot} can be rewritten in terms of calculable Monte Carlo estimators,
\begin{equation}
	\label{eq:Adotmc}
	\dot{A} = \frac{1}{\Delta t}\frac{1}{V} \sum k_{\nu} \epsilon \ell
\end{equation}
The emission rate can be simplified using the Planck-mean opacity, $k_P$, which leads to 
\begin{equation}
	\label{eq:Edotkp}
	\dot{E} = 4\pi  k_P  B(T_d) 
	             =  4\pi k_P \frac{\sigma T_d^4}{\pi} 
\end{equation}
where $B$ is the frequency-integrated Planck function and $\sigma$ is the Stefan--Boltzmann constant. Setting this equal to the absorption rate gives the dust temperature
\begin{equation}
	\label{eq:TfromAE}
	T_d = \left( \frac{ \dot{A} }{ 4 \sigma k_P} \right)^{1/4}
\end{equation}

To find the gas temperature of each cell, we calculate the solution which gives the heating rate equal to the total cooling rate. Gas is heated via ionization of neutral H and He. For species $X^i$ at ionization state $i$, the heating rate is given by
\begin{equation}
	\label{eq:ionheating}
	\begin{split}
	G(X^i) &= n(X^i) \int_{\nu_I}^\infty \frac{a\subnu{}(X^i) 4 \pi J\subnu{} }{h\nu} (h\nu - h\nu_I)  \dif \nu  \\
	            &=  n(X^i) \frac{1}{V\Delta t} \sum_{\nu_I}^\infty \frac{a\subnu{}(X^i) \epsilon \ell}{h\nu} (h\nu - h\nu_I)
	\end{split}
\end{equation}
Sources of gas cooling are free--free radiation, recombination lines from H and He, and collisionally excited forbidden lines from metals; \cref{tab:metalabundance} shows the species included in our model.

The collisional gas-dust heat exchange rate per unit volume is taken from \citet{hollenbach1979},
\begin{equation}
	\label{eq:gasgraincool}
	\Gamma_\textrm{gas-dust} = 2 f n_H n_d \sigma_d v_p  k_B (T_g - T_d)
\end{equation}
where $n_d$, $\sigma_d$, $T_d$ are the number density, cross-section and temperature of dust grains, $v_p$ is the thermal speed of protons at the gas temperature $T_g$, and $f$ is a factor which depends on the ionization state and gas temperature. 

\subsection{Radiation pressure}
\label{sec:radpressure}
We use the momentum-transfer scheme described by \citet{harries2015} to calculate the radiation pressure in each cell. As photon packets interact with a cell, they exchange momenta such that the net change in momentum between the packet leaving and entering the cell is
\begin{equation}
	\label{eq:radmomentum}
	\Delta \mathbfit{p} = \frac{\epsilon_i}{c} \Delta \hat{\mathbfit{u}}
\end{equation}
where $\hat{\mathbfit{u}}$ is the unit vector in the direction of travel. The radiative force per unit volume is then 
\begin{equation}
	\label{eq:radforce}
	 \mathbfit{f}_\textrm{rad} = \frac{\Delta \mathbfit{p}}{\Delta t V} 
\end{equation}
which is added onto the momentum equation during the hydrodynamics step.

\subsection{FUV interstellar radiation field}
The far-ultraviolet (FUV) flux $G_0$ in units of the Habing flux \citep{habing1968} is calculated in all grid cells at all time steps using
\begin{equation}
	\label{eq:g0}
	\begin{split}
	\frac{G_0}{\SI{1.63e-3}{\erg\per\s\per\cm\squared}} &=  \int_{\lambda=\SI{2400}{\angstrom}}^{\SI{912}{\angstrom}} 4 \pi J\subnu{} \dif \nu \\
	&= \frac{1}{\Delta t V} \sum_{\lambda=\SI{2400}{\angstrom}}^{\SI{912}{\angstrom}} \epsilon \ell 
	\end{split}
\end{equation}
\citep{osterbrock06}. 

\subsection{MC estimator smoothing}
In order to increase the efficiency of the Monte-Carlo estimators for the radiation field we use a scheme where each of the previous estimates are weighted according to how many time steps ago they occurred and then averaged. The weighting for each estimate of the radiation field is given by 
\begin{equation}
	\label{eq:MCweights}
	 w_i = \exp(\frac{k \Delta t}{t_\textrm{rad}})
\end{equation}
where $k$ is the number of time steps ago the estimate was made, $\Delta t$ is the time step of the simulation and $t_\textrm{rad}$ is the radiation timescale. For a sufficiently large number of previous estimates the total of the weights can be approximated to the infinite sum
\begin{align}
	\sum_{k=0}^{\infty} e^{-ak} &= \frac{e^a}{e^a -1} \\
    a &\equiv \frac{\Delta t}{t_\textrm{rad}} \\
    \label{eq:MCweightSum}
    f_\textrm{sum} &= \sum_{k=0}^{\infty} \left( f_k e^{-ak} \right) 
\end{align}
Using this formulation of the weights allows us to retain all the information for all the previous radiation history as a single value (equation \ref{eq:MCweightSum}). In order to calculate the weighted radiation value for the next timestep from the instant estimate of that time step and the weighted sum from the previous time step, we can use the fact all the weights from the previous timestep are simply multiplied by $e^{-a}$ to give the weights for the next time step. This allows us to calculate the new weighted radiation value using 
\begin{equation}
	\label{eq:MCcalc}
	f_{n,\textrm{weighted}} = \left( f_n + e^{-a} f_\textrm{sum} \right) \frac{e^a -1}{e^a} 
\end{equation}
Once this has been done $f_\textrm{sum}$ is set to the new value of $f_n + e^{-a} f_\textrm{sum}$ for the next time step. For the value of $f_\textrm{sum}$ at $t=0$ we assume the radiation field has been static for a long time so that $f_\textrm{sum} = f_0 \frac{e^a}{e^a - 1}$.

This method gives an improved estimate of the radiation field by drawing on more information at the cost of introducing some time lag into the radiation field as it changes.

\subsection{Hydrodynamics}
\label{sec:hydro}
A hydrodynamics step takes place after each radiation step. \torus{} solves the Eulerian equations of mass conservation and momentum conservation,
\begin{align}
	\label{eq:massconservation}
	\pdv{\rho}{t} + \div{(\rho \mathbfit{u})} &= 0  \\
	\label{eq:momentumconservation}
	\pdv{\rho \mathbfit{u}}{t} + \div{((\rho \mathbfit{u}) \mathbfit{u})} &= 
	    - \grad{P} - \rho \grad{\phi} + \mathbfit{f}_\textrm{rad}
\end{align}
using a finite-volume, explicit-differencing method, assuming the hydrodynamics evolves isothermally. The equations are solved using an operator splitting method. A V-cycling multigrid method is used to solve Poisson's equation,
\begin{equation}
	\label{eq:poisson}
	\nabla^2 \phi = 4 \pi G \rho
\end{equation}
with Dirichlet boundary conditions based on a multipole expansion. Mass is allowed to flow out through the boundaries but not in.

\subsection{Stars}
Stars are represented by moving Lagrangian sink particles as implemented by \citet{harries2015}, based on \citet{federrath2010}. This implementation was used by \citet{harries2017} to model the formation of a single massive star on sub-parsec scales. In our simulation, since we do not resolve down to these scales, we do not initiate sink accretion, but we still include the gravitational forces. Stars begin on the zero-age main sequence (ZAMS) and follow stellar evolution tracks by \citet{schaller1992} between 0.8 to \SI{120}{\msol}. We use the tracks of two masses, $M_1,M_2$, such that the initial stellar mass lies between $M_1 < M_{*,\textrm{ini}} < M_2$. We interpolate to the find the new mass at the current age in the $M_1$ track, repeat for the $M_2$ track, then interpolate between the two resulting masses, yielding the final new mass. We follow the same procedure for the luminosity, effective temperature and radius. Spectra of O-stars follow the \textsc{ostar2002} grid of models calculated using \textsc{tlusty} by \citet{lanz2003}, while later-type stars follow the models of \citet{kurucz1991}. 

\section{Initial conditions}
\label{sec:initialconditions}
We carry out our simulations on a 3D grid with a uniform resolution of $256^3$. This resolution is chosen such that the point of complete gas dispersal is reached within a reasonable computing wall time; we have also tested models at lower resolution and results converge at $256^3$. 
The initial condition is a spherical cloud with a uniform-density inner core extending to half the sphere radius, with the outer half going as $r^{-1.5}$. The density outside the sphere is 1 per cent of the density at the sphere edge. These conditions are similar to other cluster simulations such as \citet{krumholz2011a} and \citet{howard2016}. The sphere has a total mass $M_s= \SI{1000}{\msol}$, radius $R_s = \SI{2.66}{\pc}$, and 
mean surface density $\Sigma_s = \SI{0.01}{\g\per\cm\squared}$. The size of the grid is approximately six times the sphere radius at $\SI{15.5}{\pc}$, giving a resolution of \SI{0.06}{\pc} per cell. We impose the same random Gaussian turbulent velocity field as \citet{bate2002}, with a power spectrum $P(k)\propto k^{-4}$ for wavenumber $k$, such that the kinetic energy equals the gravitational potential energy, i.e. the virial parameter $\alpha_\textrm{vir} \equiv 2 E_\textrm{kin}/E_\textrm{grav} = 2$. 

\begin{figure}
	\includegraphics[width=\columnwidth]{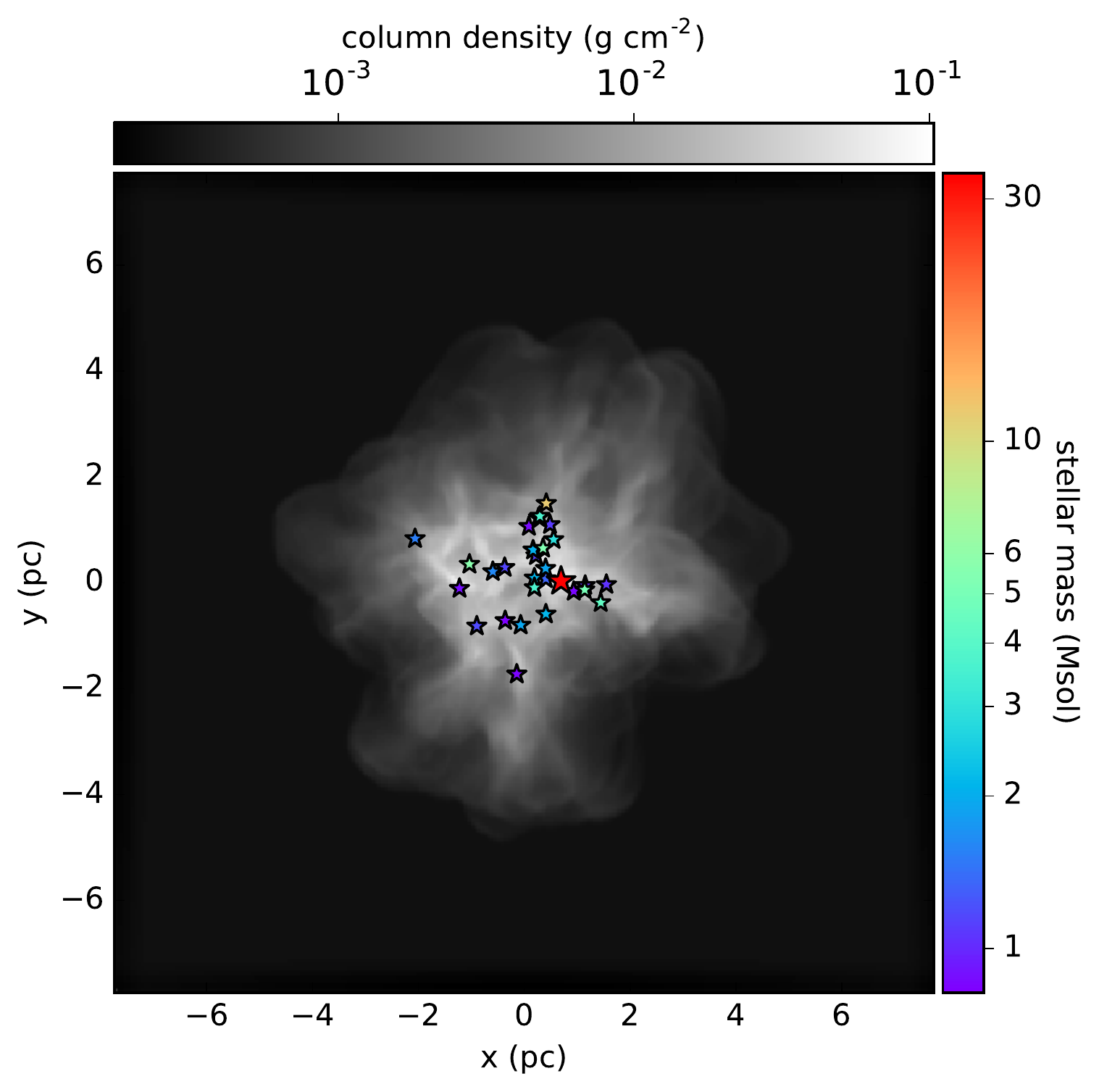}
    \caption{Positions of stars at the onset of feedback, with stellar mass in colour scale, overlaid on column density in greyscale (both are logarithmic). The most massive star is \SI{33.7}{\msol} in red. The second highest is \SI{11.3}{\msol}. The third is \SI{5.7}{\msol}. The least massive is \SI{0.82}{\msol}.}
    \label{fig:starburst}
\end{figure}

We evolve the clouds under gravity and turbulence without stars up to 0.75 $\mean{t_\textrm{ff}}$, where $\mean{t_\textrm{ff}} = \SI{2.17}{\mega\yr}$ is the average free-fall time associated with a sphere of uniform density $\rho = 3M_s/4 \pi R_s^3$; this is when \citet{krumholz2011a} see a SFE of 10 per cent. At this time we randomly sample stars from a \citet{salpeter1955} initial mass function such that the cumulative stellar mass is 10 per cent of the cloud mass (\SI{100}{\msol}) and at least one massive star is present. The most massive star (\SI{33.7}{\msol}) is placed in the cloud's most massive clump. The other 28 stars (the next most massive being \SI{11}{\msol}) are placed according to a probability density function assuming a star formation rate $\dot{M}(\mathbfit{r})$ at some position $\mathbfit{r}$; that is, $p(\mathbfit{r}) \propto \dot{M}(\mathbfit{r}) \propto \rho(\mathbfit{r})/t_\textrm{ff} \propto \rho(\mathbfit{r})^{1.5}$. The initial radius, luminosity, effective temperature and ionizing photon production rate of the massive star are listed in \cref{tab:starparameters}.
The initial distribution of stars is shown in \cref{fig:starburst} overlaid on top of column density. At this point, we switch on the radiation field and evolve the simulation until all the mass leaves the volume.

\begin{figure}
	\includegraphics[width=\columnwidth]{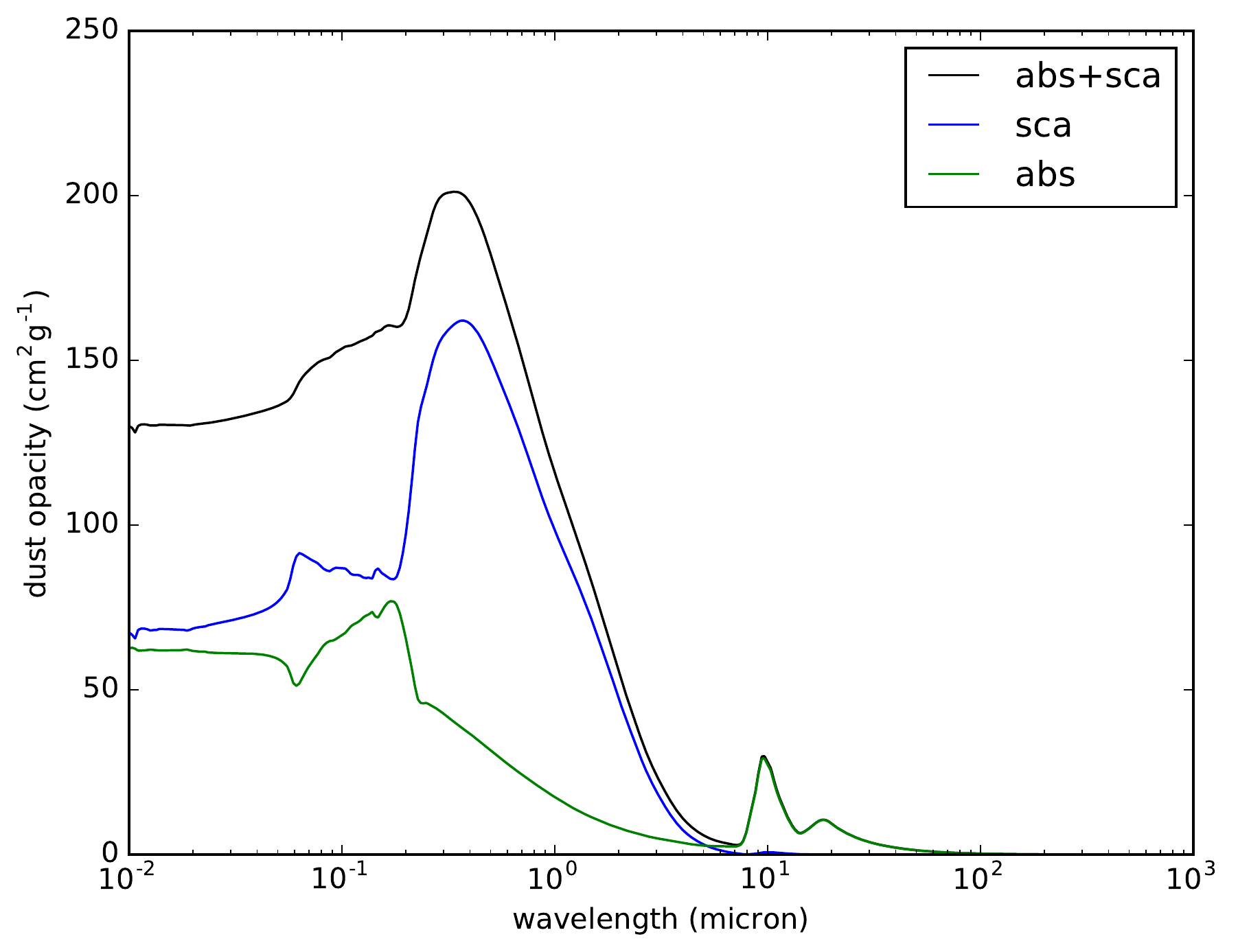}
    \caption{Dust opacity per unit mass as a function of wavelength showing scattering, absorption, and total opacities.}
    \label{fig:opacity}
\end{figure}

Elemental abundances are listed in \cref{tab:metalabundance}, using the same values as \citet{haworth2015}. We include the first few ionized states of each element, with ionization fractions calculated using the photoionization equilibrium \cref{eq:ionbalancemc}. The total abundance of each element remains constant. We also include silicate dust grains using properties from \citet{draine1984} -- \cref{fig:opacity} shows the dust absorption, scattering, and total opacities as a function of wavelength. We use a constant dust-to-gas ratio of 0.01 and follow a standard ISM power-law density distribution \citep{mathis1977}
\begin{equation}
	\label{eq:dustdistribution}
    n(a) \propto a^{-q}
\end{equation}
using grain sizes $a$ between 0.1 to \SI{1}{\micron} and a power-law index $q$ of 3.5, giving a median grain size of \SI{0.12}{\micron}. 


\begin{table}
	\centering
	\caption{Total abundance of each element and the ionized states included in the simulation.}
	\label{tab:metalabundance}
	\begin{tabular}{lcl} 
		\hline
		Element & $\log_{10}$(abundance) (rel. to H) & Ionized states \\
		\hline
		Hydrogen & 0 & I--II \\
		Helium & -1 & I--III \\
		Carbon & -3.66 & I--IV \\
		Nitrogen & -4.40 & I--III \\
		Oxygen & --3.48 & I--III \\
		Neon & -4.30 & I--III \\
		Sulphur & -5.05 & I--IV \\
		\hline
	\end{tabular}
\end{table}

\begin{table}
	\centering
	\caption{Initial parameters of the massive star.}
	\label{tab:starparameters}
	\begin{tabular}{lc} 
		\hline
		Parameter & Value \\
		\hline
		Mass & \SI{33.7}{\msol}  \\
		Luminosity & \SI{1.49e5}{\lsol}  \\
		Radius & \SI{7.59}{\rsol}  \\
		Effective temperature & \SI{41189}{K} \\
		Ionizing flux ($h\nu \geq \SI{13.6}{\eV}$) & \SI{7.36e48}{\per\s} \\
		
		\hline
	\end{tabular}
\end{table}
%%%%%%%%%%%%%%%%%%%%%%%%%%%%
%%%%%%%%% RESULTS
%%%%%%%%%%%%%%%%%%%%%%%%%%%%
\section{Results}
In this section we present the results from a model with both photoionization and radiation pressure feedback, along with a model with just photoionization (i.e. $\textbfit{f}_\textrm{rad}$ in \cref{eq:momentumconservation} is set to zero).

\label{sec:results}

\label{sec:rhdresults}
\begin{figure*}
	\includegraphics[width=\textwidth]{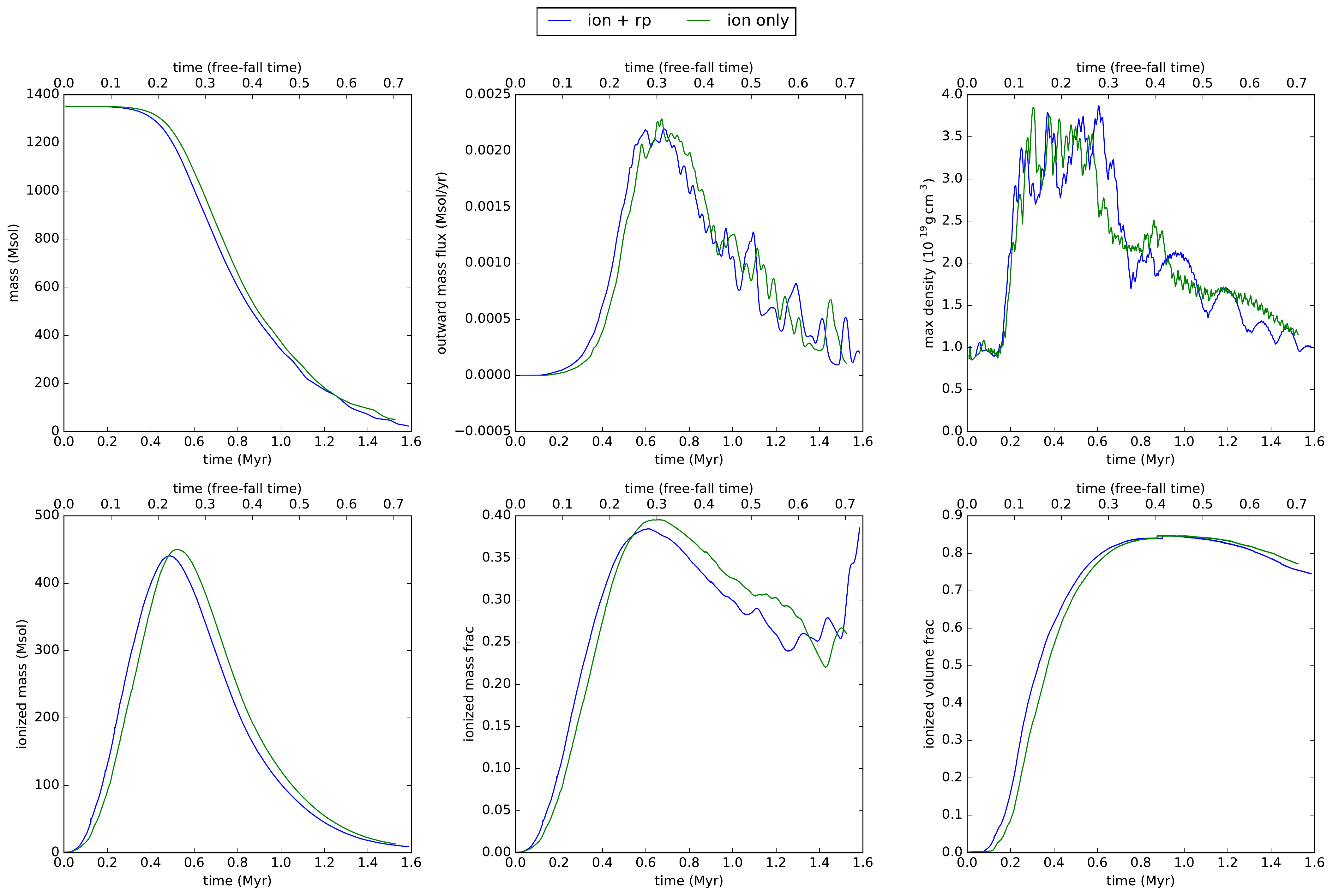}
    \caption{Bulk grid properties as a function of time, showing total mass, mass flux off the grid, maximum mass volume density, ionized mass, ionized mass fraction, and ionized volume fraction. The blue line is the model with ionization and radiation pressure; the green line is only ionization. $t=0$ corresponds to the onset of feedback.}
    \label{fig:plotoftotals}
\end{figure*}

\begin{figure*}
	\includegraphics[width=\textwidth]{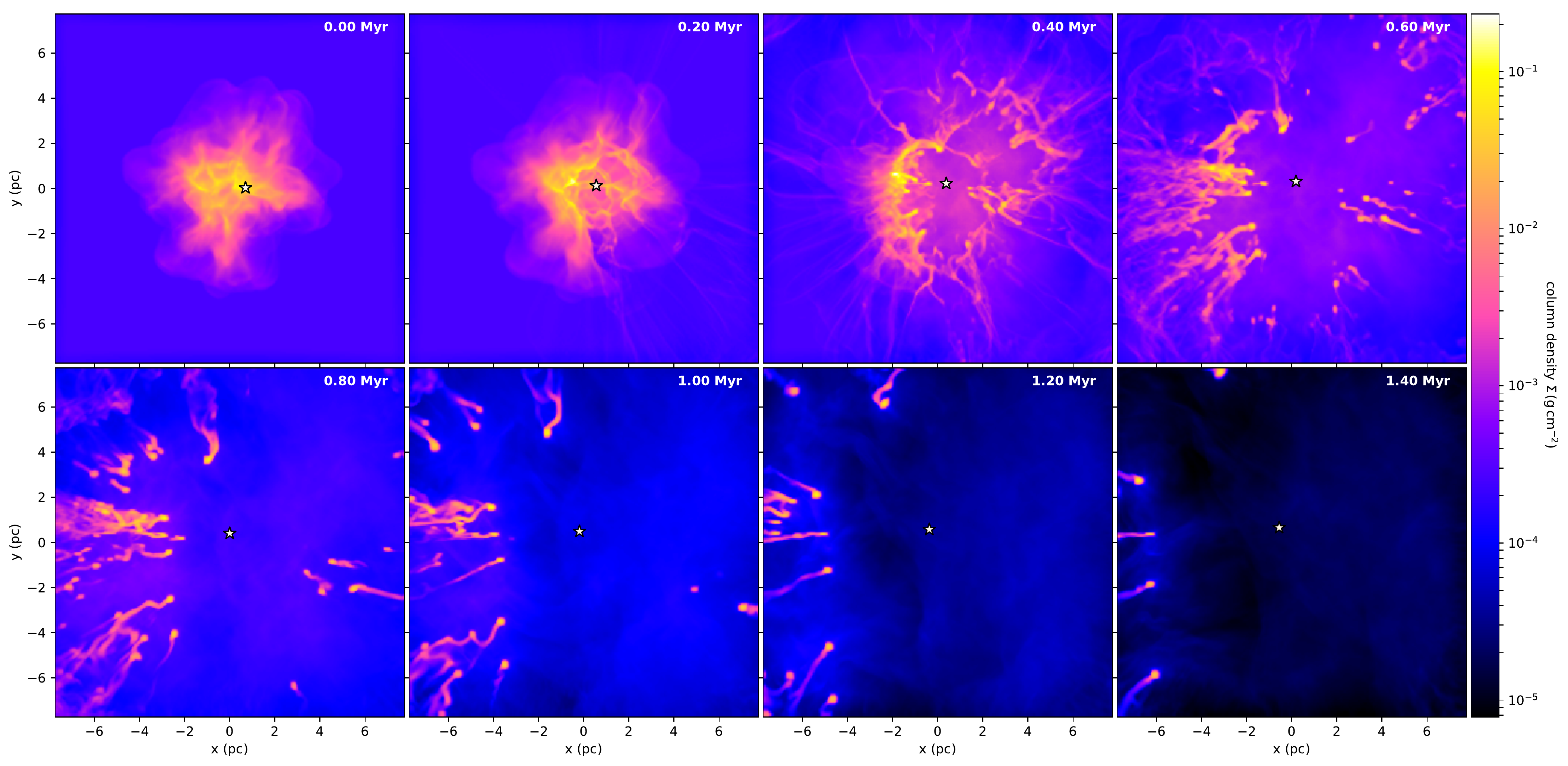}
    \caption{Column density integrated along the z-axis at 0.2 Myr intervals in the combined feedback model. Each frame is \SI{15.5}{pc} a side. The \SI{34}{\msol} star is indicated with a point near the centre of each frame. $t=0$ corresponds to the onset of feedback.}
    \label{fig:columndensity}
\end{figure*}

\subsection{Bulk grid properties}
\label{sec:bulk}
In \cref{fig:plotoftotals} we plot as a function of time the total mass on grid, mass flux off the grid, maximum density, ionized mass and mass fraction, and ionized volume fraction. \Cref{fig:columndensity} shows column density at \SI{0.2}{\mega\yr} snapshots.

The peak mass flux off the grid is \SI{2e-3}{\msol\per\yr}, reached \SI{0.6}{\mega\yr} after initiating feedback -- this is the ionized sound-crossing time from the centre of the grid to the boundary. From the onset of feedback ($t=0$) to \SI{0.6}{\mega\yr}, the mass flow is steady, with low-density ionized gas streaming through the channels carved out of the high-density filaments by the expanding ionization front. The envelope of diffuse gas outside of the initial cloud is also pushed outwards by the expanding gas. The right half of the cluster (in \cref{fig:columndensity}) is dispersed more quickly than the left half, which contains higher-density structures and is therefore more resistant to destruction and dispersal. They remain on the grid and close to their initial positions, but the ionization front creates holes in low-density areas, and curves around high-density areas, creating clumps with tails pointing away from the ionizing star. These objects move radially away from the massive star due to the rocket effect caused by photoevaporation along their star-side edges \citep[as in e.g.][]{bertoldi1990,mellema1998}. They approach the edge of the volume over the course of the simulation with an average speed $\approx \SI{6}{\km\per\s}$. After \SI{0.6}{\mega\yr} the overall mass flux begins to decrease but contains spikes corresponding to the removal of the clumps. The size of the spikes grows with time, as the densest clumps are the last to leave the grid. By about \SI{1.6}{\mega\yr}, or $0.74\,\langle t_\textrm{ff} \rangle$, all the mass has left the $(\SI{15.5}{\pc})^3$ volume. 

The third panel of \cref{fig:plotoftotals} shows the maximum density in \si{\g\per\cm\cubed} as a function of time. This peaks at just under \SI{4e-19}{\g\per\cm\cubed} at \SI{0.6}{\mega\yr}. Between 0.2 and \SI{0.6}{\mega\yr} are when the densities become highest, as the expansion of the \HII{} region drives material together. In the first \SI{0.2}{\mega\yr}, as gas gets ionized, the dense core containing the massive star expands spherically outwards, colliding with another set of dense filaments nearby ($\approx \SI{e-19}{\g\per\cm\cubed}$; at $(x,y) \approx (-1,0)$ pc in \cref{fig:columndensity}, which shows column density). Between 0.2 and \SI{0.4}{\mega\yr} the outflung material sweeps across the filament, with the densest areas remaining somewhat stationary while the lower-density gas is carried along with the flow. During this process, the filament is compressed and material that is initially perpendicular to the expanding shell is broken up and carved into pillars oriented parallel to the flow, ending up as tails behind spherical cores (pointing radially away from the ionizing star). Compression of the filament causes the maximum density to increase, reaching its highest value of \SI{4e-19}{\g\per\cm\cubed} at \SI{0.6}{\mega\yr}. Once the expanding material has passed through and the pillars are formed, the densest cores are more exposed to the stellar radiation field and there is less collisional compression -- photoevaporation removes material from the clumps, parts of the pillars break off into separate chunks, and the maximum density falls.

The fourth and fifth panels of \cref{fig:plotoftotals} show the total ionized mass and the ionized mass fraction, respectively. The highest value of ionized mass is \SI{440}{\msol} at \SI{0.5}{\mega\yr} (36 per cent of the total mass). The peak ionized mass fraction is reached \SI{0.1}{\mega\yr} later, still just under 40 per cent of the total mass. This is despite 85 per cent of the volume being ionized (sixth panel of \cref{fig:plotoftotals}), showing that most of the mass remains in small, dense clumps which resist ionization.

The photoionization-only model is mostly similar to the model with both photoionization and radiation pressure, but there are a few minor differences. The bulk effect of feedback is delayed in the photoionization-only model, with the peak in the ionized mass fraction being reached \SI{0.03}{\mega\yr} later. The later breakout of ionization results in a slightly different distribution of gas, as it has had more time to evolve under gravity and turbulence, so the total amount of gas being ionized is affected (very marginally) -- the peak ionized mass fraction is about 2 per cent higher. The removal of gas from the grid also occurs with the same delay. Overall, the differences are negligible.

\subsection{Morphology}
\begin{figure}
	\includegraphics[width=\columnwidth]{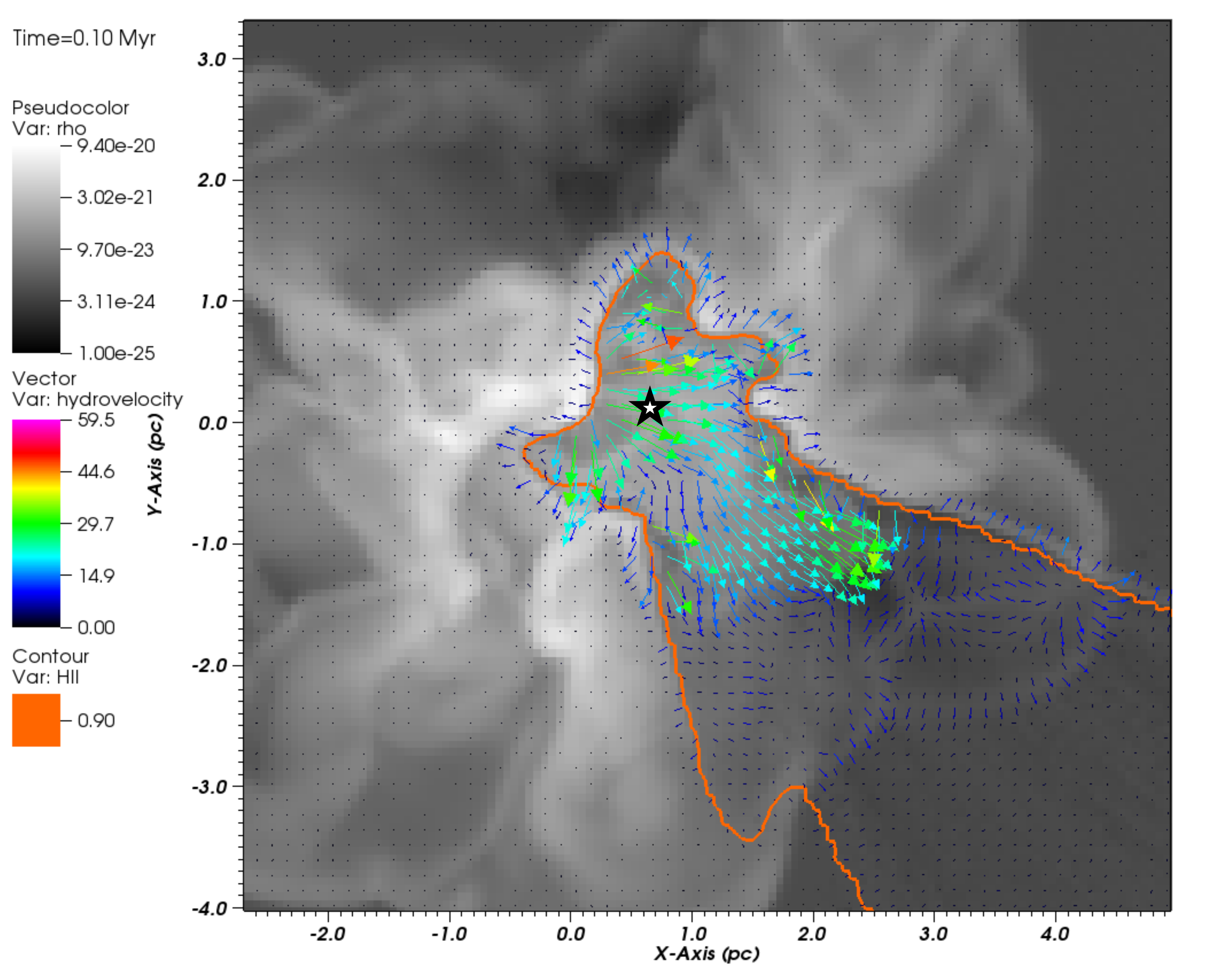}
    \caption{Zoomed-in slice in the xy plane at 0.1 Myr showing mass density in greyscale between \SI{e-25} and \SI{9.4e-20}{\g\per\cm\cubed}; vectors with size and colour corresponding to velocity between 0 and \SI{60}{\kilo\m\per\s}; and a contour where the ionization fraction is 0.9. The \SI{33.7}{\msol} star is at position $(x,y) = (0.6, 0.1)$ pc. This model includes ionization and radiation pressure.}
    \label{fig:velocityslice}
\end{figure}

Snapshots of column density in \Cref{fig:columndensity} show how the destruction of the cloud proceeds via the expansion of ionized gas over the course of \SI{1.6}{\mega\yr} in the model containing both photoionization and radiation pressure. \Cref{fig:velocityslice} shows a 2D slice of mass density and velocity vectors at \SI{0.1}{\mega\yr}. Most of the \HII{} region is confined at \SI{0.1}{\mega\yr} by high-density filaments, but a champagne flow breaks out through the low-density region on the edge of the cloud on the opposite side. At the boundaries of the \HII{} region, the gas travels outwards at the ionized sound speed (approximately \SI{12}{\km\per\s}), while the photoevaporation of gas on the inside of the boundary leads to outflows moving inwards and then out through the champagne flow at velocities of 20 to \SI{30}{\km\per\s}, with a few cells around \SI{40}{\km\per\s}. 

\begin{figure}
	\includegraphics[width=\columnwidth]{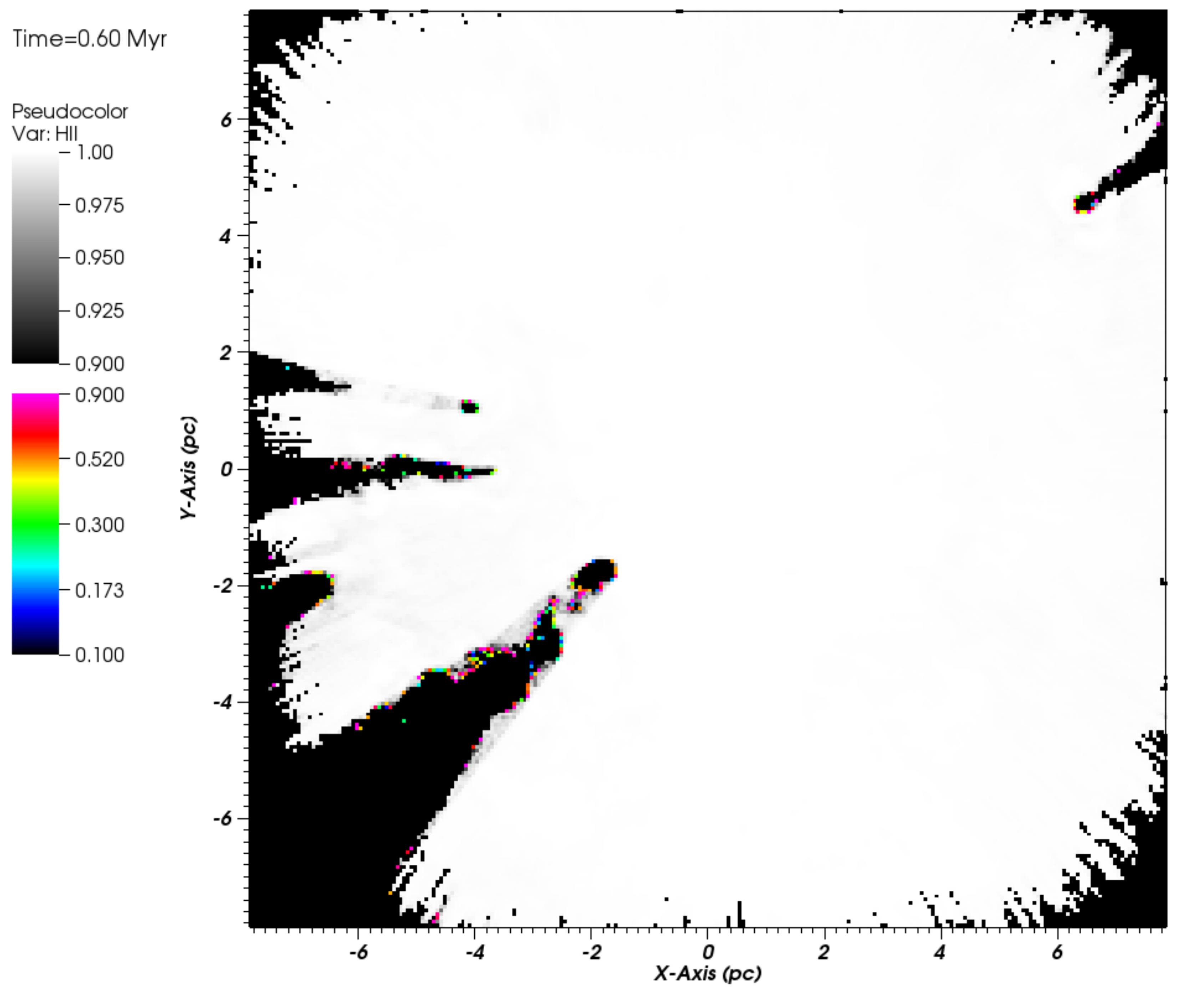}
    \caption{2D slice in the xy plane at \SI{0.6}{\mega\yr} showing the hydrogen ionization fraction. The linear greyscale is the fraction between 0.9 and 1 (fully ionized gas). The logarithmic colour scale is the fraction between 0.1 and 0.9 (essentially fully neutral to partially ionized). The densest clumps and pillars are neutral but have partially ionized edges. The interior is fully ionized.}
    \label{fig:hiifrac}
\end{figure}

The \HII{} region is able to break through some parts of the filament (e.g. near $(x,y)=(0.8, 1.4)$ pc in \cref{fig:velocityslice}), while curving around nodes such as $(x,y)=(0.1, 0.1)$ pc which resist photoionization and are carved into globules and pillars by the expanding ionization front. The densest clumps shield material that is downwind of the ionizing source as seen in \cref{fig:hiifrac}, which shows the ionization fraction of hydrogen at \SI{0.6}{\mega\yr}. Some of the shielding is only partial, for example at $(-4, 1.5)$ pc of that frame, as the diffuse radiation field ionizes gas behind the clump but to a lesser degree. This highlights the importance of including the diffuse field in RHD models.

\label{sec:columndensitypdf}
\begin{figure}
	\includegraphics[width=\columnwidth]{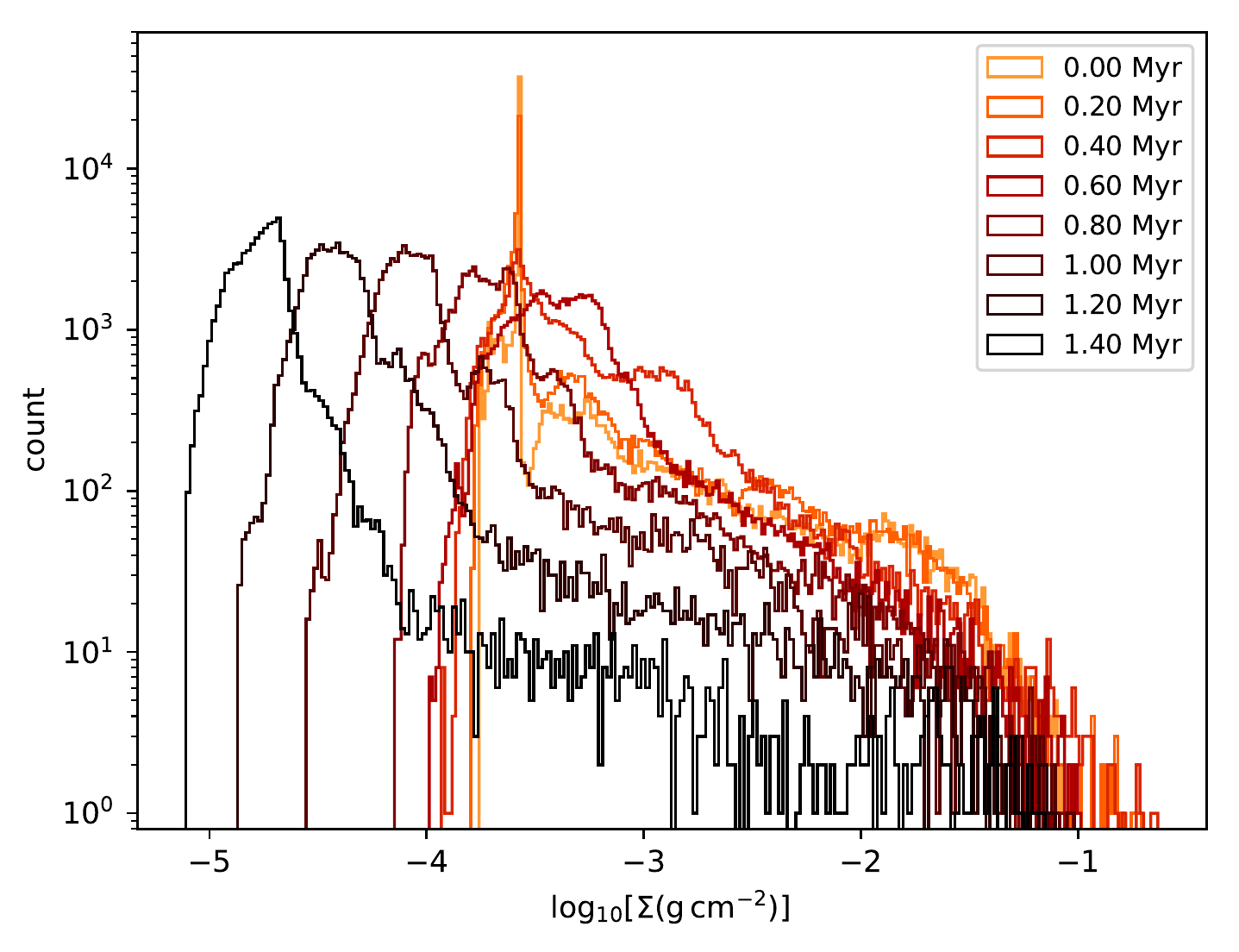}
    \caption{Column density histograms at 0.2 Myr intervals for the combined feedback model. Histograms shift towards lower densities as time progresses. The spike at $t=0$ (the onset of feedback) corresponds to the stationary, uniform-density material outside the turbulent cloud.}
    \label{fig:densitypdf}
\end{figure}

\Cref{fig:densitypdf} shows column density histograms for the combined feedback model at \SI{0.2}{\mega\yr} intervals. The spike at $t=0$ (the onset of feedback) at $\Sigma\approx\SI{3e-4}{\g\per\cm\squared}$ corresponds to the stationary gas outside around the gravoturbulent cloud, which is only perturbed when the star photoionizes it or cloud material expands into it, at which point the surrounding density increases. Overall, as the simulation evolves, the PDF shifts towards lower densities, as a result of the \HII{} region expansion and the increase in low-density, ionized material. The early high-density hump at $\approx$ a few \SI{e-2}{\g\per\cm\squared}, produced during the initial starless collapse phase, is flattened out in the first \SI{0.4}{\mega\yr}. Although higher densities are achieved up to the same period, they are not long-lived, as the maximum density is reduced after another \SI{0.2}{\mega\yr}.  

\subsection{Electron temperature and density}
\begin{figure}
	\includegraphics[width=\columnwidth]{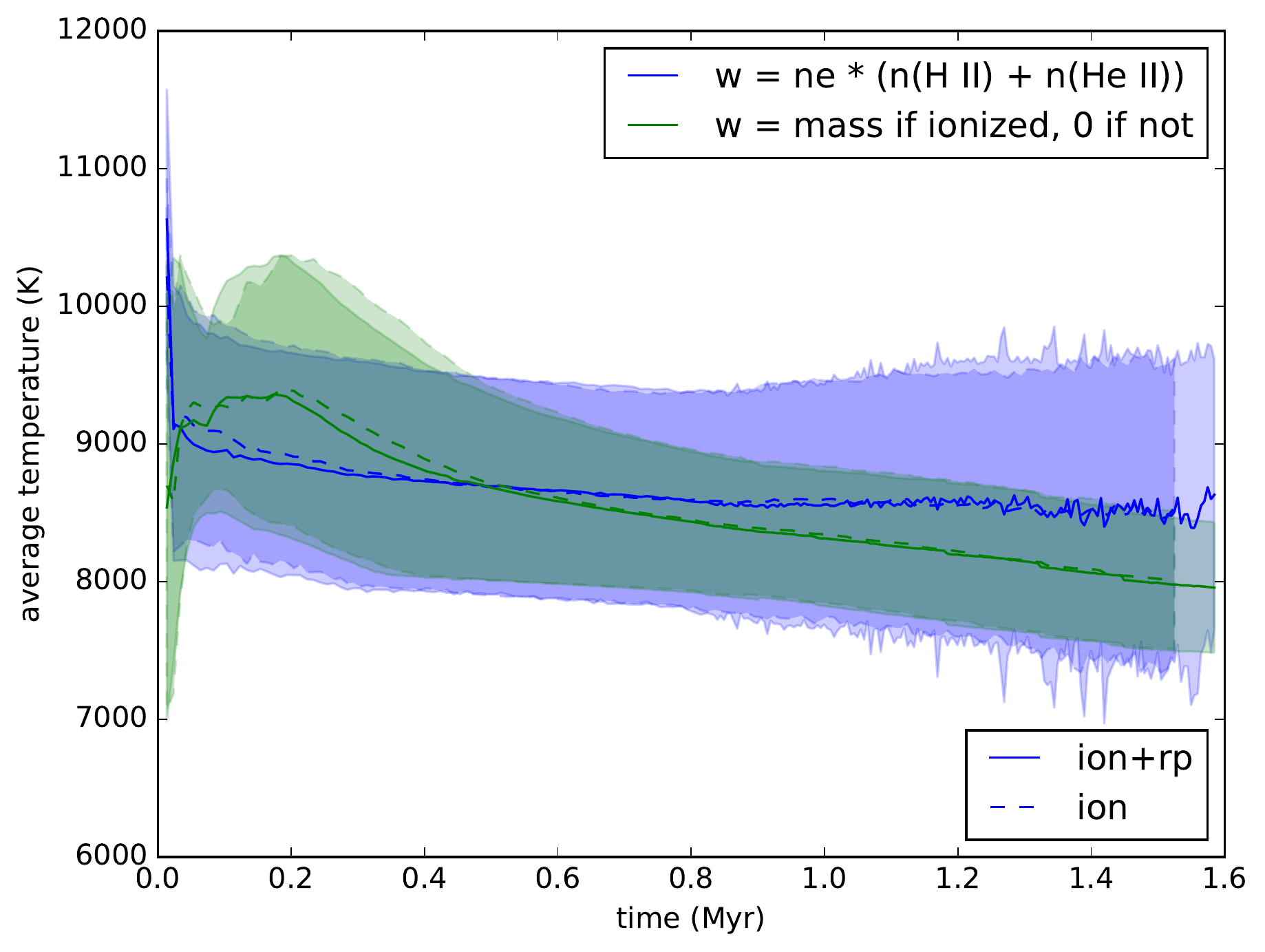}
    \caption{Volume-average gas temperature (with filled boundaries showing the standard deviation), plotted against time. The average is weighted by $w = n_e (n_{\ion{H}{II}}+ n_{\ion{He}{II}}) \approx n_e^2$ (blue), mass $w=\rho \Delta V$ if the cell is more than 90 per cent ionized or $w=0$ if less (green). Solid lines are the model with ionization and radiation pressure; dashed lines are the model with just ionization.}
    \label{fig:globaltemperature}
\end{figure}

In order to gauge the temperature of the ionized gas, we calculate weighted averages over the volume using
\begin{equation}
	\label{eq:weightedaverage}
	T_0 = \frac{ \int w T \dif{}V }{ \int w \dif{}V} =  \frac{ \sum_i w_i T_i \Delta V}{\sum_i w_i \Delta V} 
\end{equation}
where $T_i$ is the temperature of cell $i$ with volume $\Delta V$, and we consider two different weights $w$: (a) $w = n_e (n_{\ion{H}{II}}+ n_{\ion{He}{II}})$ as per \citet{rubin1968}, which is well approximated by $w \approx n_e^2$; (b) mass $w=\rho \Delta V$ if hydrogen in the cell is more than 90 per cent ionized or $w=0$ if less.  In \cref{fig:globaltemperature} we plot the volume-average temperature as a function of time.

The mass-weighted average temperature is highest at \SI{0.2}{\mega\yr} having a value \SI{9300}{K} whilst the ionization front is still largely contained inside the cloud, but then decreases to \SI{8000}{K} over the next \SI{1.4}{\mega\yr}. The $n_e^2$-average is \SI{9000}{K} over the whole duration. The standard deviation is about 10 per cent for the $n_e^2$ average and is steady until about \SI{1}{\mega\yr}, at which point the deviation rises, with greater fluctuation, towards 14 per cent. The mass-weighted average also has an standard deviation around 10 per cent, but after \SI{0.4}{\mega\yr} this drops to 6 per cent and is much more steady than the electron density-weighted average. This is due to the high-mass globules and filaments which are neutral (and hence don't contribute to either average) but whose edges are partially ionized (so they do contribute to the electron density-weighted average but not the mass-weighted average). This is visualised in \cref{fig:hiifrac} where the greyscale denotes cells which are more than 90 per cent ionized, and the colour scale shows cells which are less. Since the interaction of the ionization front and the dense, neutral gas changes relatively quickly over time, the spatial extent and degree of partial ionization similarly changes, giving rise to the fluctuation in the $n_e^2$ average. 

\begin{figure}
	\includegraphics[width=\columnwidth]{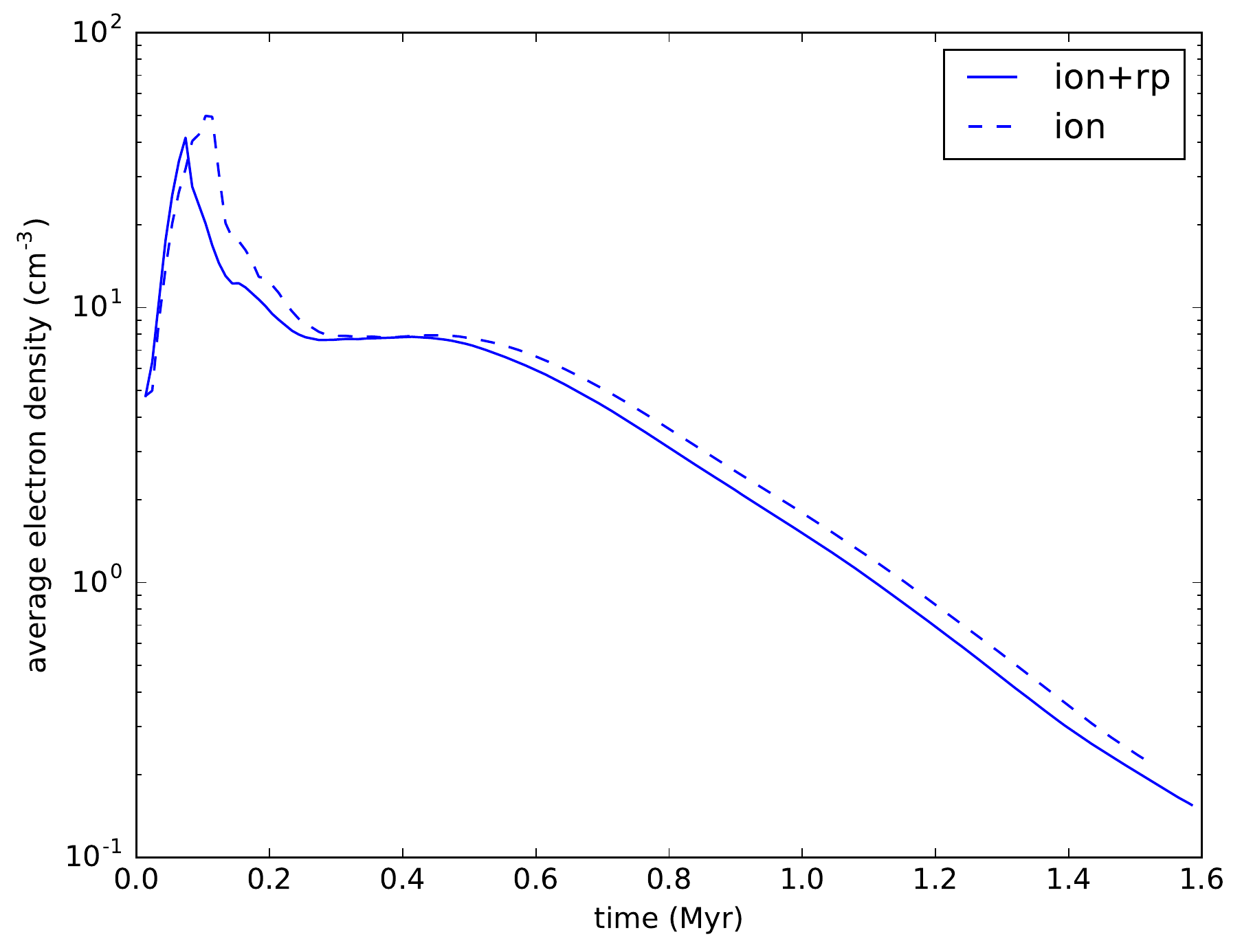}
    \caption{Volume-average electron density $n_e$ in the ionized gas, plotted against time. Solid lines are the model with ionization and radiation pressure; dashed lines are the model with just ionization.}
    \label{fig:globalne}
\end{figure}

\cref{fig:globalne} shows the (unweighted) volume-average electron density in gas which is more than 90 per cent ionized. This reaches a maximum of \SI{30}{\per\cm\cubed} at $\approx\SI{0.1}{\mega\yr}$ before dropping down to \SI{8}{\per\cm\cubed} by \SI{0.2}{\mega\yr}. It remains at this value until \SI{0.5}{\mega\yr}. As the gas flows off the grid over the next Myr the density decreases once again, reaching \SI{0.2}{\per\cm\cubed} by the end of the simulation. 
 

\subsection{FUV interstellar radiation field}
\label{sec:g0}

\begin{figure}
	\includegraphics[width=\columnwidth]{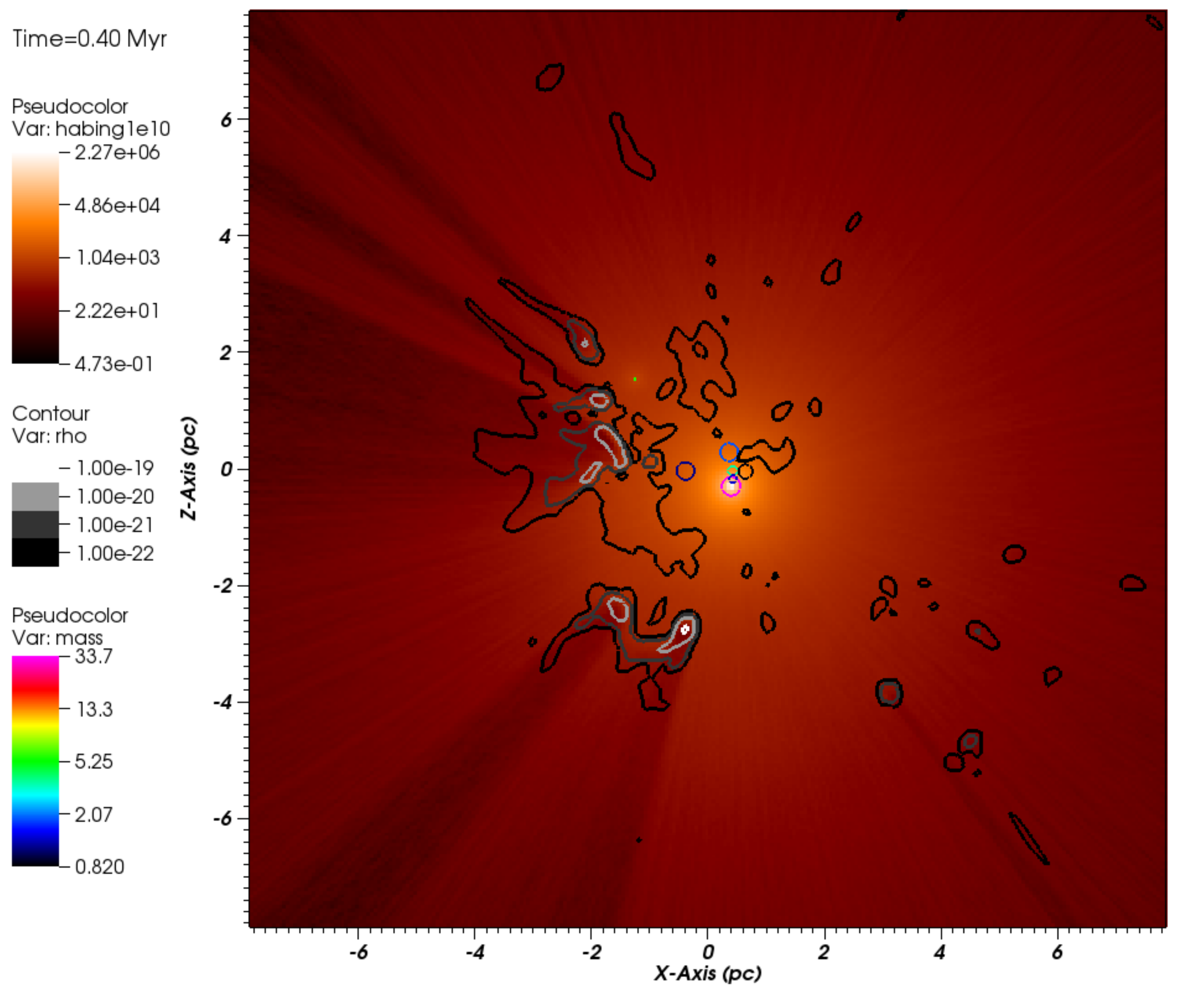}
    \caption{Interstellar FUV flux $G_0$ in units of the Habing flux (\SI{1.63e-3}{\erg\per\s\per\cm\squared}). This is a slice in the xz plane at \SI{0.4}{\mega\yr}, taken through the position of the most massive star (pink circle, \SI{34}{\msol}). Stars are plotted as 3D surfaces with a radius of 2.5 grid cells, therefore intersections of the slice with star surfaces result in circles (rainbow colour scale, with the size corresponding to the proximity to the slice). Grayscale contours show mass volume density at levels of $10^{-19}, 10^{-20}, 10^{-21}, 10^{-22}$~\si{\g\per\cm\cubed}.}
    \label{fig:habingfluxSlice}
\end{figure}
\begin{figure}
	\includegraphics[width=\columnwidth]{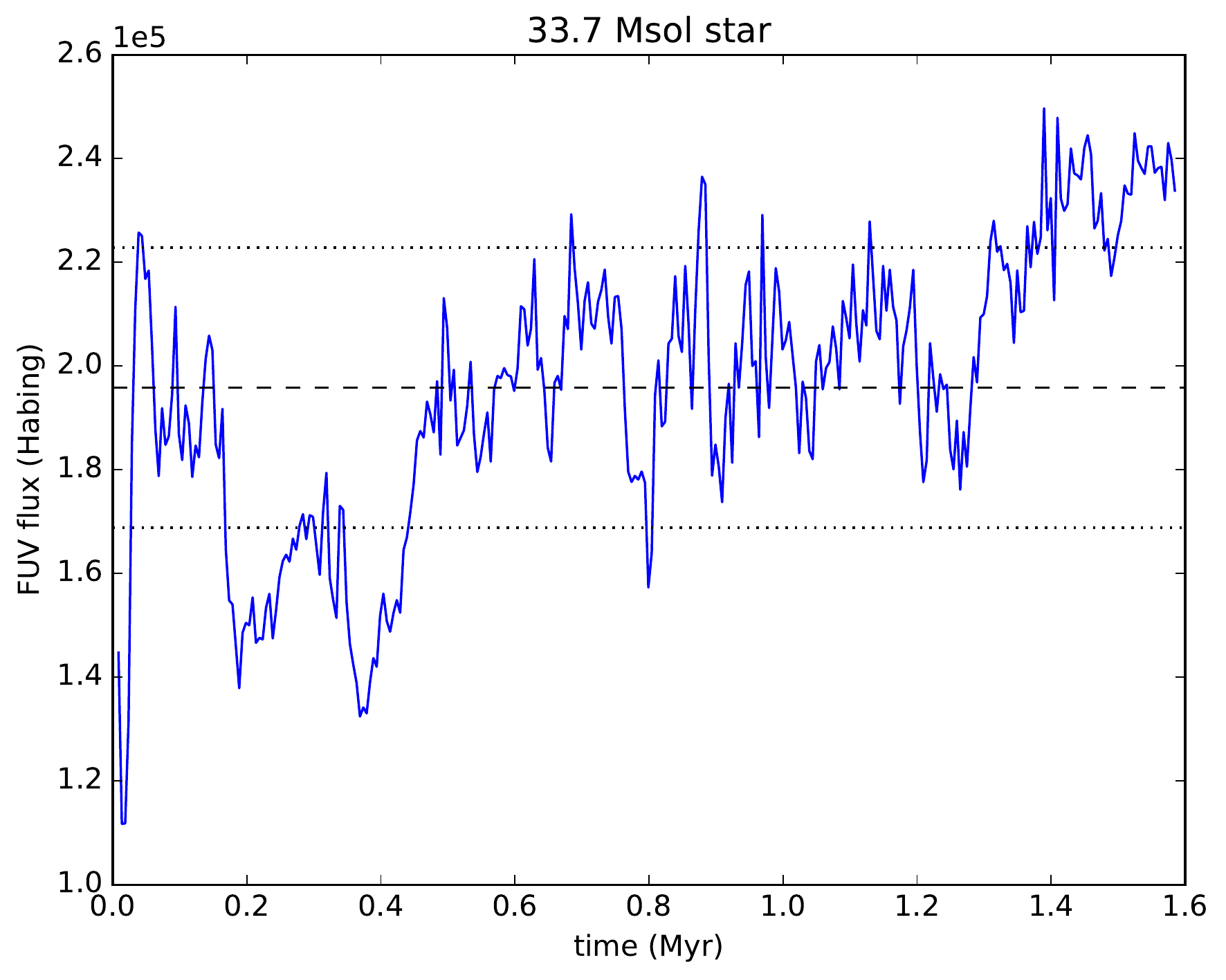}
    \caption{Interstellar FUV flux $G_0$ in units of the Habing flux (\SI{1.63e-3}{\erg\per\s\per\cm\squared}), at the location of the most massive star, as a function of time. This is the mass-weighted average within a radius of 2.5 grid cells (\SI{0.15}{\pc}) around the star. The horizontal dashed black line shows the time-average flux and the dotted lines show the standard deviation.}
    \label{fig:habingfluxStar}
\end{figure}
\begin{figure}
	\includegraphics[width=\columnwidth]{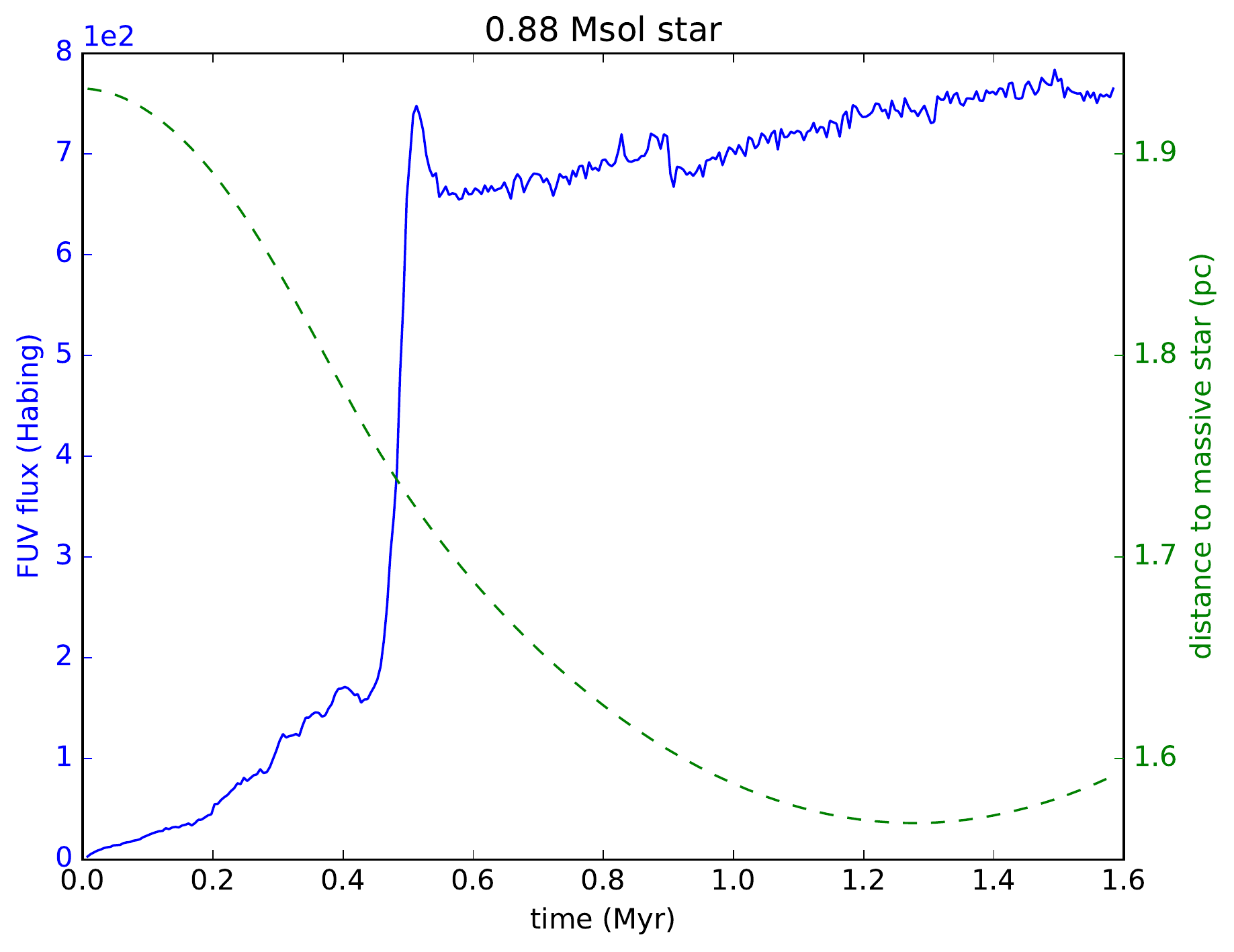}
    \caption{Left axis, solid blue line -- interstellar FUV flux $G_0$ in units of the Habing flux (\SI{1.63e-3}{\erg\per\s\per\cm\squared}), at the location of a \SI{0.88}{\msol} star, as a function of time. This is the mass-weighted average within a radius of 2.5 grid cells (\SI{0.15}{\pc}) around the star.  Right axis, dashed green line -- distance to the most massive star (\SI{33.7}{\msol}) in pc. The rapid increase in flux at \SI{0.5}{\mega\yr} is caused by the sudden exposure to the radiation field of the most massive star as material gets removed from the smaller star's surroundings.}
    \label{fig:habingfluxStar25}
\end{figure}

\Cref{fig:habingfluxSlice} shows a representative 2D slice of $G_0$, the FUV flux (\cref{eq:g0}), in the plane of the \SI{34}{\msol} star, showing the variation with distance at $t=\SI{0.4}{\mega\yr}$. The flux decreases as $r^{-2} \textrm{e}^{-\tau}$ due to geometrical dilution and the optical depth $\tau$, as in \citet{bisbas2015}, with shadowed cones caused by dense, dusty clumps which shield the downstream material from the stellar radiation field. The maximum value is of the order \SI{e6}{} in the cell containing the star, with values of order \SI{e5}{} in adjacent cells. In \cref{fig:habingfluxStar}, we plot as function of time the average $G_0$ inside a radius of 2.5 cells (\SI{0.15}{\pc}) around the star, weighted by mass. Averaging over time, this is \SI{2.0(3)e5}{} in units of the Habing flux. Fluctuations arise from the star moving between cells which have different densities and therefore different optical depths. The inner Orion Nebula is observed to have values around \SI{4e5}{} \citep{osterbrock06} which is comparable in magnitude to the region we model here.

We include a similar plot for a \SI{0.88}{\msol} star. This is shown in \cref{fig:habingfluxStar25} alongside the time-varying distance to the massive star. Between 0 and \SI{0.5}{\mega\yr}, $G_0$ rises steadily to a value of \SI{2e2}, before jumping up by more than a factor of 3.5 within  $\approx \SI{50}{\kilo\yr}$. This occurs because it is located inside a pillar which is being photoevaporated and pushed away from the ionizing source, decreasing the column density between the two stars; once it becomes completely exposed to the radiation field of the brightest source, the flux rapidly rises. After this point, $G_0$ remains relatively level up to the end of the simulation as the intervening material is diffuse, with a slight increase as the stars approach each other. We save a detailed investigation of the fluxes around the other 27 stars for a future study.

%%%%%%%%%%%%%%%%%%%%%%%%%%%%
%%%%%%%%% DISCUSSION
%%%%%%%%%%%%%%%%%%%%%%%%%%%%
\section{Discussion of the RHD model}
\label{sec:discussion}
\subsection{Dynamics}
The differences between the model with both ionization and radiation pressure, and the model with just ionization, are negligible, leading to the conclusion that photoionization is a more important process than radiation pressure for the dispersal of gas in the conditions presented here. According to \citet{draine2011} the latter process is expected to have a more significant role in higher-density clouds, e.g. $n>\SI{100}{\per\cm\cubed}$ with a star having the same Lyman continuum photon flux as our model ($\approx \SI{e49}{\per\s}$).

Gas dispersal is more pronounced than in simulations by \citet{rogers2013}, who modelled stellar winds and supernovae feedback in a cloud with similar density but slightly higher mass (\SI{3240}{\msol}), containing three massive stars between 28 to \SI{35}{\msol}. During the main sequence phase, the mass flux does not exceed \SI{5e-4}{\msol\per\yr}, or a quarter of the peak flux in our model. Although their grid is twice the size of ours, meaning material must travel further to leave the grid, the mass flux is relatively uniform over the first \SI{4}{\mega\yr}, with fluxes only increasing after the evolved Wolf-Rayet and supernova phases; fluxes higher than \SI{2e-3}{\msol\per\yr} are only achieved after this point. This implies photoionization is a more efficient feedback mechanism for dispersing clouds than stellar winds, and is more comparable to SNe, lending support to \citet{matzner2002} who concluded that \HII{} region expansion is the dominant source of feedback in GMCs. \citet{rogers2013} found that dense gas was largely unaffected by feedback, with winds dispersing through low-density channels from the initial conditions. The resilience of dense gas was also found by \citet{dale2012}. This is borne out by our model as well.

The early hydrodynamical models of champagne flows in 1D and 2D by e.g. \citet{tenorio-tagle1979}, \citet{bodenheimer1979} and \citet{yorke1989} -- of an ionizing O star located next to the interior boundary of a molecular cloud -- show the same characteristics as our 3D model presented here (see \cref{fig:velocityslice}): the ionized cloud gas escapes as a champagne flow at \SI{30}{\km\per\s} while the diffuse ionized gas outside of the cloud expands at \SI{10}{\km\per\s}. Such velocities are also observed in real \HII{} regions such as the Orion Nebula \citep[M42;][]{odell2017a}, DR 21 \citep{immer2014}, and the Hourglass in M8 \citep{chakraborty1997}. The schematic in Figure 4 of \citet{odell2009}, interpreting observations of the Orion Nebula, is remarkably similar to the champagne flow we see in \cref{fig:velocityslice}, including the supersonic gas travelling away from the massive star, as well as the ionization front stalling at dense cores, from which photoevaporated material flows back towards the ionizing source. 

The structures produced in our model are likely to be a result of the initial conditions. \citet{walch2013} found similar structures in simulations where the initial conditions were filamentary and had high fractal-dimension, whereas spherical \HII{} regions arose from conditions with lower fractal dimensions and more spherical geometries. \citet{dale2011} note that accretion flows towards deeply embedded massive stars limit the expansion of \HII{} regions, and therefore molecular clouds with such stellar distributions remain relatively undisturbed by feedback. On the other hand, when massive stars are located closer to the edge of molecular clouds, such as in the Orion Nebula, their \HII{} regions may disperse gas effectively via champagne flows \citep{henney2005}. In our model, the massive star is located deep within the cloud but it is still able to blow out a champagne flow through nearby low-density channels. Once the flow breaks out, the expanding \HII{} region is able to disrupt and disperse the rest of the cloud -- even if the starting position was dense and more resistant. 

One of the limitations of our model is that stars are placed in a particular location instead of being formed self-consistently via accretion; the star is positioned fully-formed in a node between several filaments after the initial self-gravitating, turbulent evolution of the cloud. Therefore we also do not model the evolution of an ultracompact \HII{} region alongside the growth of the star and the possible feedback effects this may incur. However, models by \citet{peters2010} of massive protostars at the sub-parsec scale show that ultracompact \HII{} regions `flicker' while the star is still accreting and only grow to substantial sizes after mass reservoirs are depleted. Therefore, as an approximation for scales of a few to \SI{10}{pc}, a significant \HII{} region only blows out into the cloud once the star reaches its final mass, which is the stage we start with in our model. Furthermore, our calculation is informative for how the gas is displaced \textit{after} this stage, and allows us to compare with, for example, models of stellar winds by \citet{rogers2013} who place three massive stars in the centre of a turbulent medium. That said, for more comprehensive and self-consistent simulations we plan to use sink particles to self-consistently grow star- or cluster-particles, using subgrid models to compensate for the limited spatial resolution. 

\subsection{Temperature}
The volume-average ionized gas temperature is approximately \SI{9000(1000)}{K} over the course of the simulation. M42 (the Orion Nebula) has a comparable electron temperature at \SI{9200(1600)}{K}, with the fluctuation depending on the observational diagnostic \citep{odell2001}. 

\citet{haworth2015} provided a temperature parameterisation of the same thermal balance calculation as in our model, but for an \HII{} region expanding into an initially uniform-density medium. The ionized gas temperature is described by
\begin{equation}
    \label{eq:haworthTempIonfrac}
    T_i = T_n + \eta \left[1.1 \times 10^4 - 3.8 \times 10^3 \left(\frac{z}{z_0} - 0.5\right)^{0.839} - T_n\right] 
\end{equation}
where $T_n$ is a prescribed fully neutral gas temperature (e.g. \SI{10}{K} or the dust temperature), $\eta$ is the ionization fraction of hydrogen, and $z$ is the metallicity relative to the Lexington benchmark metallicity $z_0$ \citep[which we also use here;][]{ferland1995,ercolano2003,haworth2012}. \citeauthor{haworth2015} used the same gas heating and cooling rates as our model, and so it accurately matches our volume-average temperature (\SI{9000}{K}; \cref{fig:globaltemperature}). (The two models used slightly different dust size distributions, so the gas-grain heat exchange term would be different; however, in the ionized gas, the rates of ionization heating and metal cooling would dominate over the gas-grain exchange rate.) This equation may be useful for those looking to use a simplified temperature scheme to account for the same thermal balance terms as our calculation, provided the ionization fraction is already known. However, it does not take into account the scatter in temperature which is about 10 per cent. 

\subsection{FUV interstellar radiation field}
In \cref{sec:g0} we showed how the FUV flux reaching a \SI{0.88}{\msol} star increases rapidly as gas is removed from its surroundings. Photons at these wavelengths (912 to \SI{2400}{\angstrom}) cause photolelectric heating of dust and photodissociation of H$_2$ \citep{draine1978,osterbrock06}. Protoplanetary disks, or proplyds, around such stars in real star-forming regions are therefore stripped of material as the thermal pressure increases, resulting in a photoevaporative wind blowing from their outer layers \citep{odell1993,kim2016a}. This is not negligible, as external irradiation from other stars can be greater than internal irradiation by many orders of magnitude \citep{bruderer2012}. Disk models typically include some external source of flux that remains constant in time \citep[e.g.][]{haworth2016}, but this is not representative of real clusters where stars move around and gas is displaced -- a proplyd may see a radiation field that switches `on' or `off' depending on the intervening gas dynamics. This may in turn affect proplyd dispersal rates. In a future study, we intend to characterise the time-varying nature of the FUV flux around all 28 stars in our simulation.

%%%%%%%%%%%%%%%%%%%%%%%%%%%%
%%%%%%%%% SYNTHETIC OBSERVATIONS
%%%%%%%%%%%%%%%%%%%%%%%%%%%%
\section{Synthetic observations}
\label{sec:syntheticobservations}
Synthetic observations are produced using the temperatures, densities, dust properties, elemental abundances and ionization fractions that were calculated and evolved during the RHD model -- they were not modified with any post-processing. In this section, we analyse the model with both photoionization and radiation pressure. 

\subsection{Recombination and forbidden lines}
\label{sec:recombinationforbidden}
\begin{figure*}
	\includegraphics[width=\textwidth]{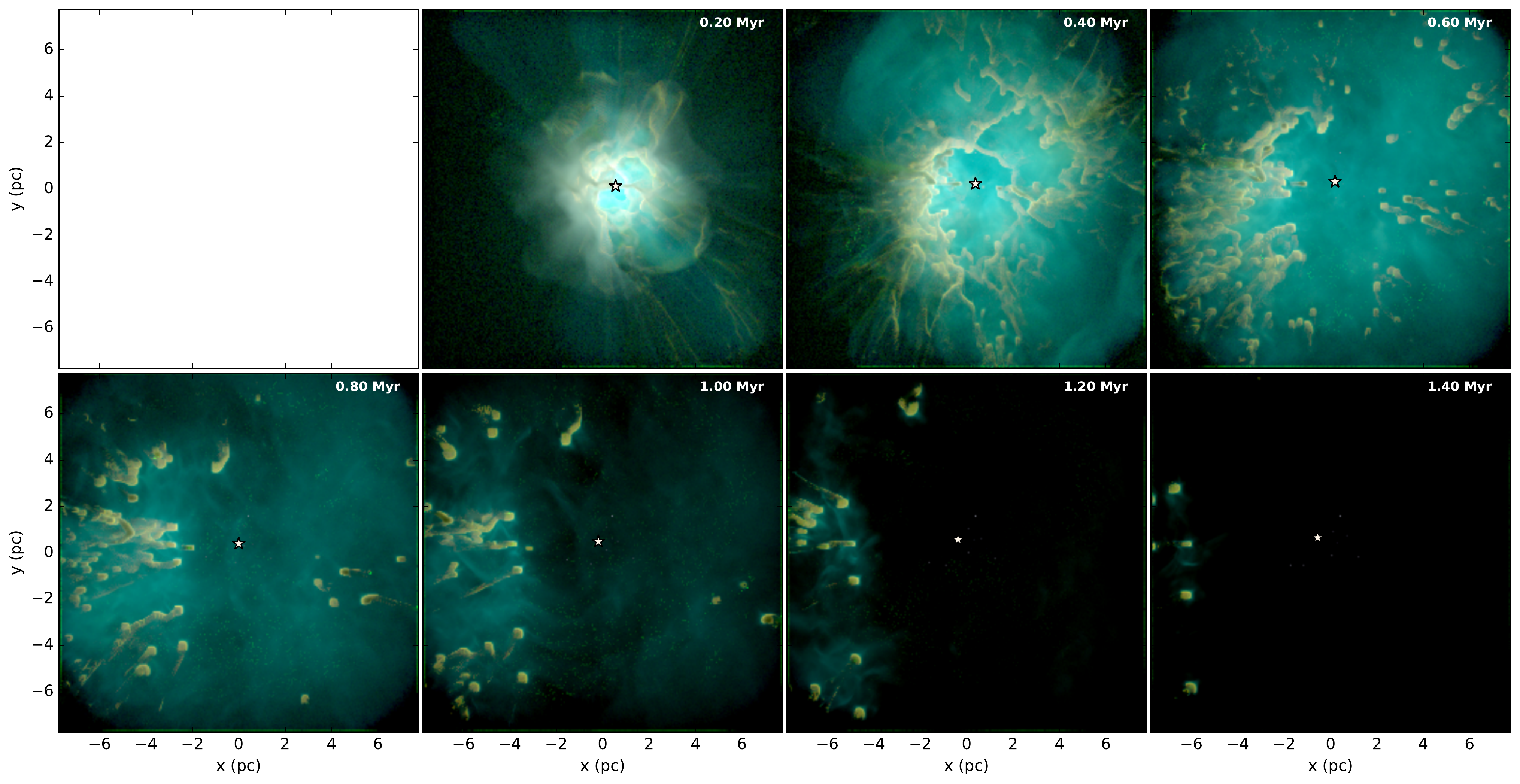}
    \caption{Three-colour images using synthetic observations of [\ion{S}{II}] \SI{6731}{\angstrom} (red), H$\alpha$ (green), and [\ion{O}{III}] \SI{5007}{\angstrom} (blue), at 0.2 Myr intervals (corresponding to the column density maps in \cref{fig:columndensity}). The \SI{34}{\msol} star is indicated by a point near the centre of each frame.}
    \label{fig:rgb}
\end{figure*}

We produce synthetic observations of the hydrogen recombination line H$\alpha$ at \SI{6563}{\angstrom}, and collisionally excited metal forbidden lines of [\ion{S}{II}] at \SI{6731}{\angstrom} and of [\ion{O}{III}] at \SI{5007}{\angstrom}. \cref{fig:rgb} shows a three-colour composite at \SI{0.2}{\mega\yr} intervals corresponding to the column density snapshots in \cref{fig:columndensity}. Each line is scaled up to have the same minimum and maximum. As a representative example of the range of original surface brightnesses, at \SI{0.6}{\mega\yr} the [\ion{S}{II}] brightness varies between 0.03 and \SI{9000}{\mega\jy\per\str}; H$\alpha$ lies between 0.3 and \SI{3e4}{\mega\jy\per\str}; and [\ion{O}{III}] has values between 0.1 and \SI{e4}{\mega\jy\per\str}. All three lines are strongest around clumps of high density, where cooling is more efficient, with the [\ion{S}{II}] line showing the greatest difference in brightness between diffuse gas and dense gas. This causes the brown colour in the composite. From \SI{0.4}{\mega\yr} onwards there are several clumps with bright-rimmed envelopes and tails pointing radially away from the massive star. These characteristics are shared with the proplyds, elephant trunks, and cometary knots seen in well-known \HII{} regions spanning multiple size scales, such as the Orion Nebula, NGC 7293 \citep[the Helix, a planetary nebula;][]{odell2000}, the Carina nebula \citep{haikala2017}, the Eagle nebula, and the Rosette cloud \citep{tremblin2013}. We save a full characterisation of masses and sizes for a future study, but values are of the order 1 to \SI{30}{\msol} and 0.1 to \SI{1}{\pc\squared}. 

\subsubsection{H$\alpha$ line luminosity}
\label{sec:halphaluminosity}
\begin{figure}
	\includegraphics[width=\columnwidth]{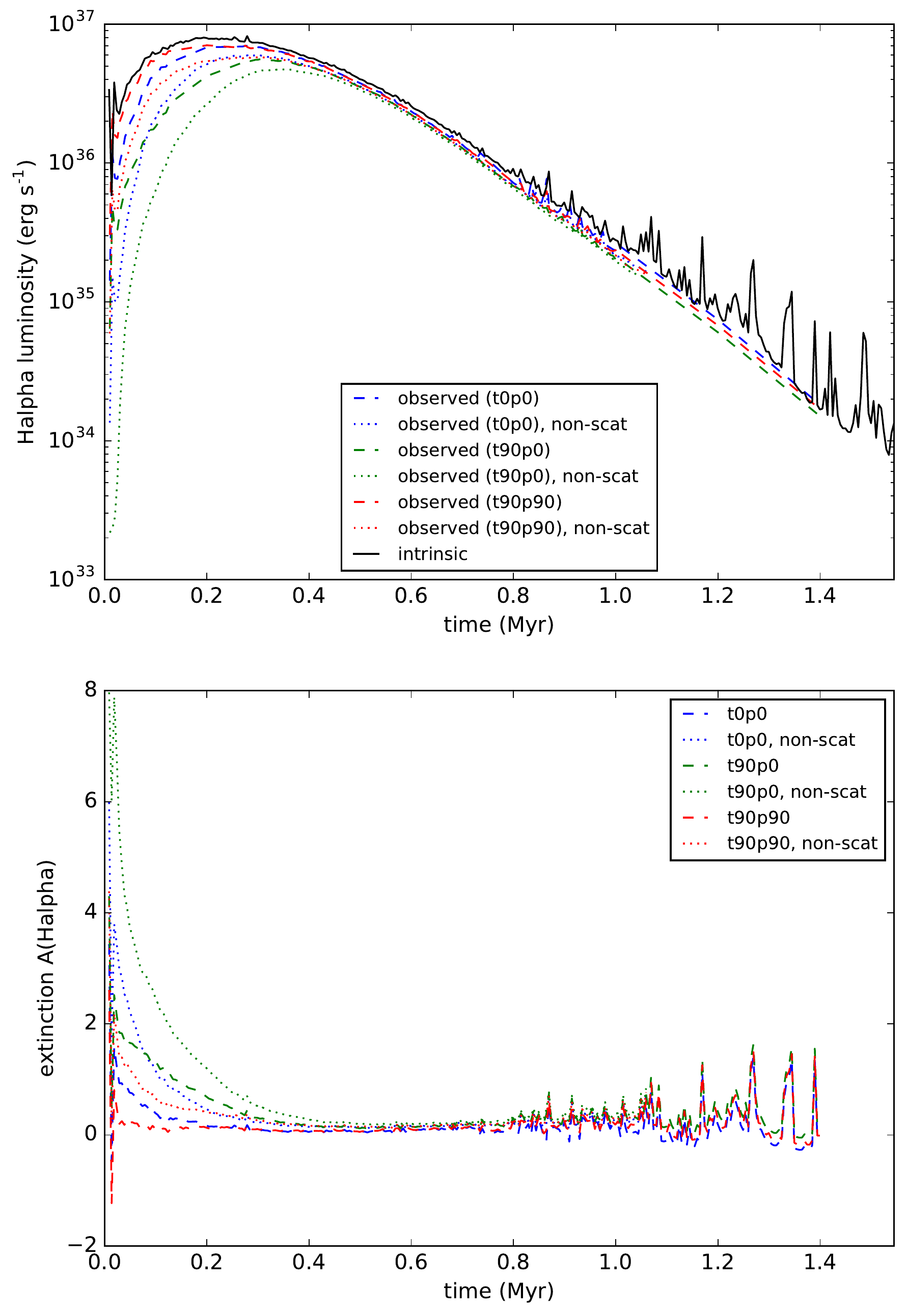}
    \caption{Top - H$\alpha$ luminosities as a function of time, with the intrinsic luminosity in solid black, and observed luminosities at viewing angles $(\theta,\phi) = (0,0), (90,0), (90,90)^\circ$ in blue, green, and red respectively. Observations without taking into account scattering from other lines of sight are in dotted lines, observations with scattering are dashed -- see \cref{sec:halphaluminosity} for a full description. Bottom - extinctions for the aforementioned parameters.}
    \label{fig:halphaluminosity}
\end{figure}
In \cref{fig:halphaluminosity} we plot the intrinsic H$\alpha$ luminosity $L_\textrm{int}$ and the observed luminosity $L_\textrm{obs}$ as a function of time, where 
\begin{equation}
	\label{eq:lintrinsic}
	 L_\textrm{int} = \int 4 \pi j\subnu{} \dif{} V
\end{equation}
and
\begin{equation}
	\label{eq:lobserved}
	 L_\textrm{obs} = 4 \pi d^2 S_{\nu}
\end{equation}
where $j\subnu{}$ is the emission coefficient, $S_{\nu}$ is the H$\alpha$ flux density, or surface brightness integrated over solid angle, and we have arbitrarily observed the model at a distance $d=\SI{400}{\pc}$ (the distance to the Orion Nebula) from three different viewing angles $(\theta,\phi) = (0,0), (90,0), (90,90)$ with the colatitude $\theta$ and azimuthal angle $\phi$ in degrees. Within the cloud itself, $L_\textrm{obs}$ is reduced along the line of sight due to absorption and scattering by dust. The extinction $A(\textrm{H}\alpha)$ in magnitudes is
  \begin{equation}
	\label{eq:extinction}
	 A(\textrm{H}\alpha) = - 2.5 \log_{10} \left( \frac{L_\textrm{obs}}{L_\textrm{int}} \right)
\end{equation}
and this is also plotted in \cref{fig:halphaluminosity}. The peak intrinsic luminosity, \SI{8e37}{\erg\per\s}, is reached at \SI{0.2}{\mega\yr} and this drops below \SI{e35}{\erg\per\s} after a \SI{}{\mega\yr}. The extinction is a few magnitudes within the first \SI{0.4}{\mega\yr} while the \HII{} region is still embedded within the cloud, after which the extinction drops to zero as the region becomes optically thin. 

Two schemes of scattering are considered. The first uses the peel-off method \citep[see][]{yusef-zadeh1984}: for photon packets which start off travelling away from the line of sight, their scattering events forcibly direct some light towards the observer (regardless of the new direction of the photon packet). This therefore adds scattered light from other lines of sight into the observed beam. Additionally, photon packets which do travel directly towards the observer may be scattered away from the observed beam. The second scattering scheme (labelled `non-scat' in \cref{fig:halphaluminosity}) only accounts for the latter effect -- scattering from other lines of sight is not included.

The extinction between different viewing angles differs by 1 to 2 magnitudes. Furthermore, the extinction with the full scattering treatment is lower by a magnitude compared to observations neglecting the peeled-off photons. This is because the emitting gas is surrounded by dense filaments of dust which has a scattering opacity peaking near \SI{6563}{\angstrom} (see \cref{fig:opacity}), and this directs light \textit{towards} the observer, partially compensating for absorption and scattering  \textit{away} from the observer. Fluctuations in the extinction after \SI{0.8}{\mega\yr} arise from small differences in ionization; this has a more pronounced effect on the luminosity at late times as the luminosity is already dim, the ionized gas is diffuse, and recombination lines are sensitive to the square of the density.

\subsection{Free--free radio observations and Lyman flux}
\label{sec:freefree}
\label{sec:nlyman}

\begin{figure}
	\includegraphics[width=\columnwidth]{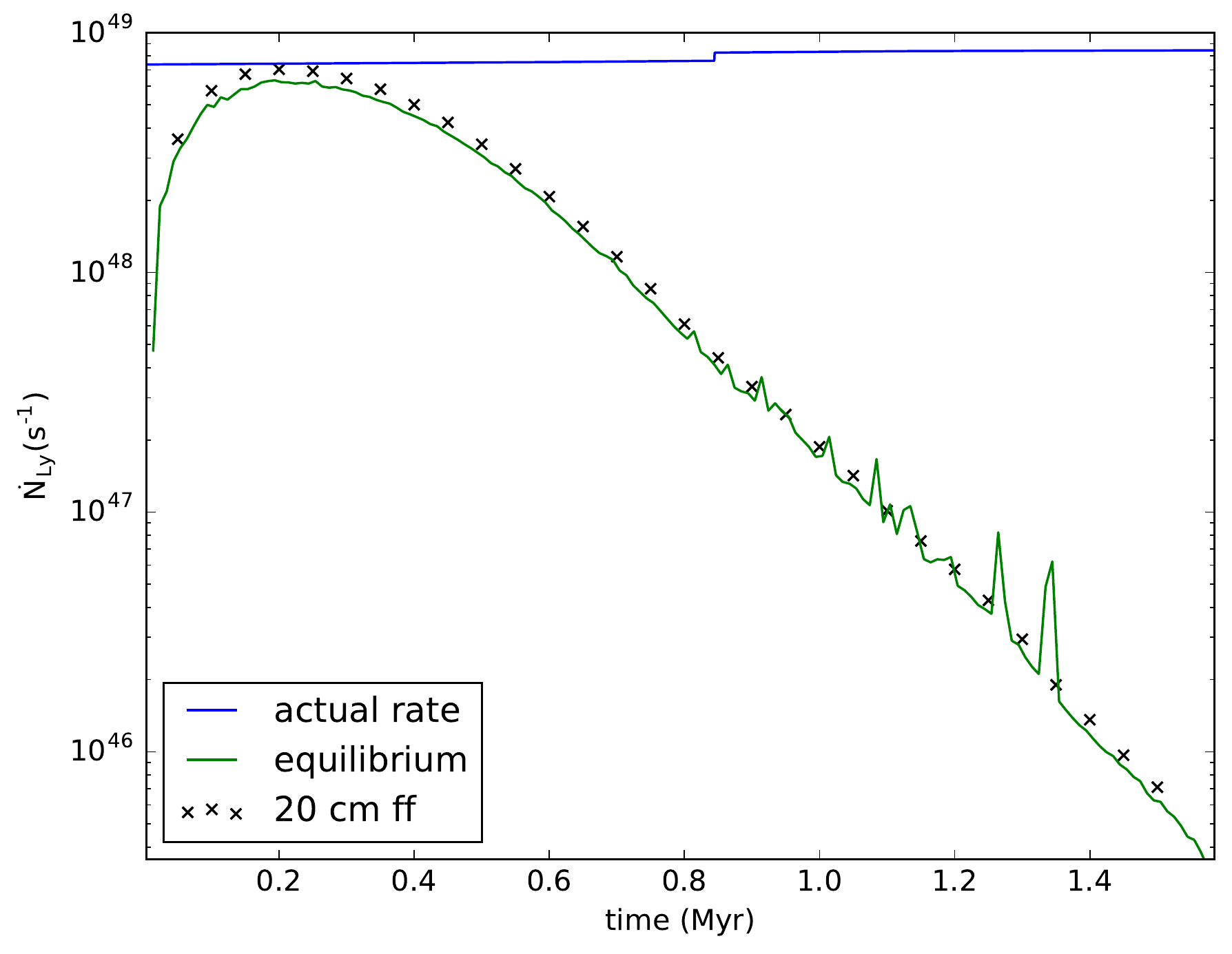}
    \caption{Total production rate of Lyman continuum photons as a function of time. Actual rate from integrating stellar spectra in blue. Rate from assuming photoionization equilibrium,  \cref{eq:nlymanequilibrium}, in green. Rate inferred from 20 cm free-free emission, \cref{eq:nlyman}, as black crosses.}
    \label{fig:nlyman}
\end{figure}

It is possible to estimate the rate of Lyman continuum photons ($\lambda < \SI{912}{\angstrom}$; $h\nu > \SI{13.6}{\eV}$) being produced in a nebula using radio observations. \citet{rubin1968} derived an expression to calculate this assuming photoionization equilibrium, which implies 
\begin{equation}
	\label{eq:nlymanequilibrium}
	\dot{N}_\textrm{Ly} \geq \int n_e  [n(\ion{H}{II}) + n(\ion{He}{II})] \alpha(T)\dif{} V
\end{equation}
where $\alpha(T)$ is the recombination rate coefficient. \citeauthor{rubin1968} uses the same $\alpha$ for H and He, and to be consistent with the literature and to test the resulting diagnostic, we do the same in our analysis. This takes the form
\begin{equation}
	\label{eq:alphab}
	\alpha(T) = 4.10 \times 10^{-10} T^{-0.8}
\end{equation}
The optical depth for free-free radiation is
\begin{equation}
 	\label{eq:tauff}
 	\tau\subnu^\textrm{ff} = 8.235 \times 10^{-2} \left( \frac{T}{\textrm{K}} \right)^{-1.35} \left( \frac{\nu}{\textrm{GHz}} \right)^{-2-\beta} \int \left( \frac{ n_e n_i }{\textrm{cm}^{-6}} \right) \frac{\dif{} s}{\textrm{pc}}  
\end{equation}
\citep{altenhoff1960,mezger1967} where $n_i = n(\ion{H}{II}) + n(\ion{He}{II})$. If $\tau\subnu^\textrm{ff} \ll 1$, the free-free surface brightness is 
\begin{equation}
 	\label{eq:brightnessff}
    I\subnu{} \approx \frac{2 k T \nu^2}{c^2} \tau\subnu^\textrm{ff}  
\end{equation}
Comparing with \cref{eq:nlymanequilibrium} leads to \citeauthor{rubin1968}'s expression for the Lyman continuum production rate
\begin{equation}
	\label{eq:nlyman}
	\dot{N}_\textrm{Ly}
	\geq 4.76 \times 10^{42} \left(\frac{\nu}{\textrm{GHz}}\right)^{\beta} \left(\frac{D}{\textrm{pc}}\right)^{2} \left(\frac{T_0}{\textrm{K}}\right)^{-0.45} \left(\frac{S\subnu{}}{\textrm{Jy}}\right) [\textrm{s}^{-1}]
\end{equation}
where $S\subnu{}$ is the surface brightness integrated over the solid angle subtended by the object observed from a distance $D$, $T_0$ is the average temperature, $\nu$ is the observation frequency, and $\beta$ is a spectral index; this depends on the free-free Gaunt factor and is usually 0.1 in the literature \citep[e.g.][]{mezger1967,rubin1968,lefloch1997,kim2017}. \Cref{eq:nlymanequilibrium,eq:nlyman} represent a lower limit in a density-bounded \HII{} region.

We apply \cref{eq:nlyman} to infer the Lyman continuum flux from synthetic observations of \SI{20}{\cm} free--free emission at \SI{0.05}{\mega\yr} intervals. We use a representative temperature of \SI{9000}{\K} as \cref{eq:nlyman} is only weakly dependent on $T_0$ and this is the average temperature in our simulation (see \cref{fig:globaltemperature}). The spectral index $\beta$ is calculated to be 0.15. We plot the results in \cref{fig:nlyman}, along with the known production rate taken from the integrated stellar spectra. We also compare this with the equilibrium \cref{eq:nlymanequilibrium}, which does not depend on any synthetic observations.

The radio measurement closely matches the result from \cref{eq:nlymanequilibrium} at all times, showing that the radio emission is accurately measured and tracks photoionization balance consistently. Before \SI{0.2}{\mega\yr}, both results underestimate the actual photon production rate by a factor of a few. \citeauthor{rubin1968}'s method assumes that Lyman continuum photons only go into ionizing the gas, neglecting the number which are absorbed by dust grains and thermally re-emitted in the infrared, reducing the number available for the gas. Since the star at this stage is deeply embedded in a node between filaments, dust absorption is not negligible. Once the \HII{} breaks out, however, the probe becomes more reliable, matching the actual emitted photon flux at \SI{0.2}{\mega\yr}. After this stage, the radio flux decreases along with the electron density, and by \SI{0.8}{\mega\yr} the measured $\dot{N}_\textrm{Ly}$ is an order of magnitude lower than the known production rate. By the end of the simulation the discrepancy is $10^3$. Mass begins to leave the grid after around \SI{0.4}{\mega\yr}, so after this stage the ionizing photons may escape from the volume; if our cloud represents a core embedded inside a larger GMC, these photons could go on to excite gas beyond the model boundary. This would account for the increasing discrepancy at later times. Since the radio method is used by observers to get spectral classifications of O-stars, this highlights the importance of knowing the full size scale of an \HII{} region -- for a limited observational field of view, the Lyman flux can be underestimated by several orders of magnitude.

\subsection{Dust emission}

\begin{figure*}
	\includegraphics[width=\textwidth]{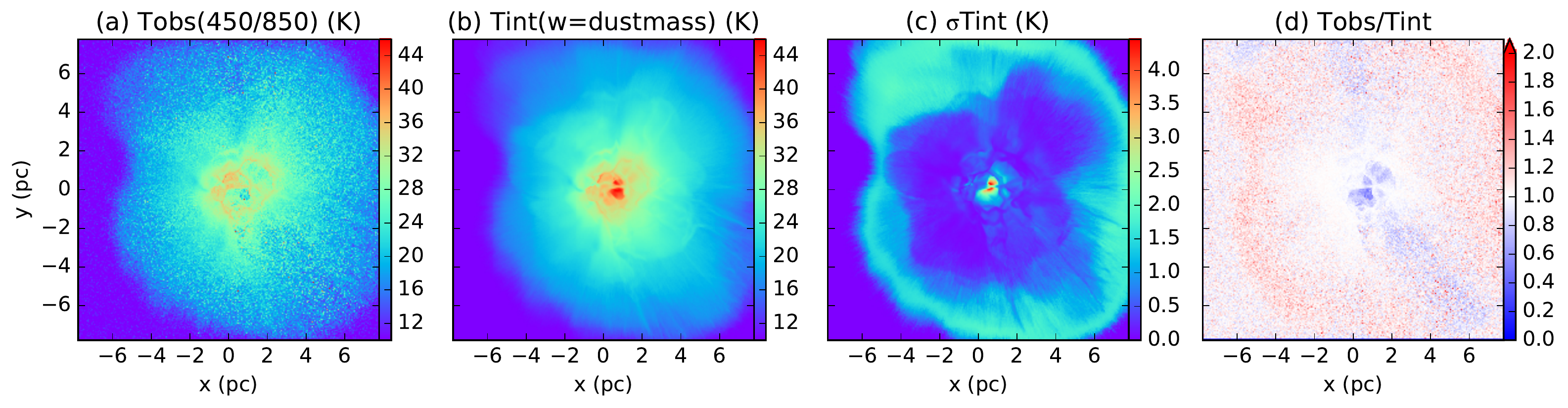}
    \caption{(a) Dust temperature extracted from 450 and \SI{850}{\micron} dust emission at \SI{0.2}{\mega\yr}. (b) Intrinsic temperature averaged along the line of sight, weighted by dust mass. (c) Standard deviation of the intrinsic temperature normalised by the average. (d) Ratio of (a) and (b).}
    \label{fig:scuba2tempsearly}
\end{figure*}
\begin{figure*}
	\includegraphics[width=\textwidth]{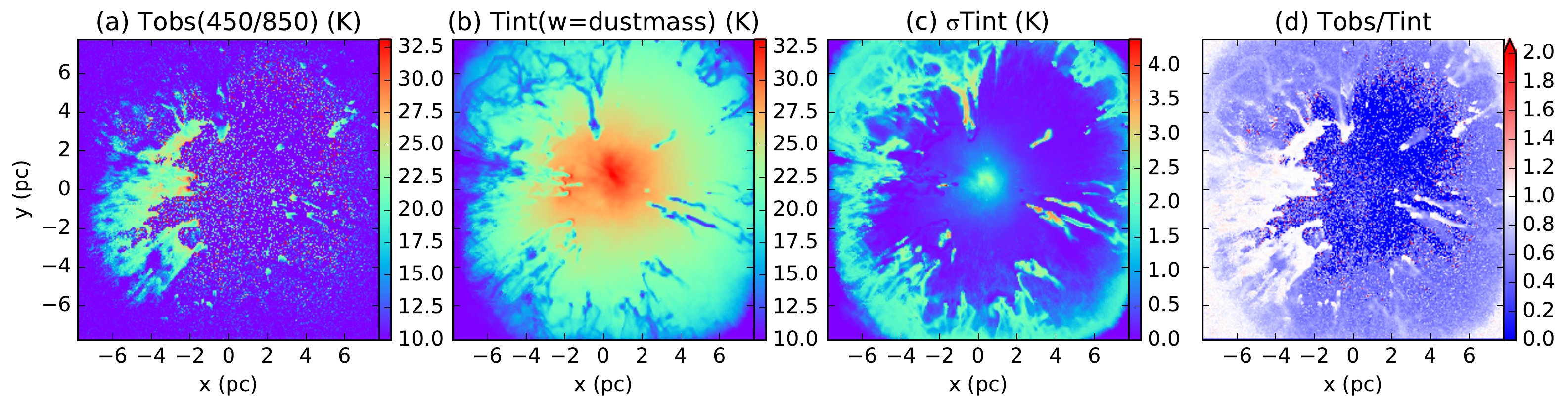}
    \caption{(a) Dust temperature extracted from 450 and \SI{850}{\micron} dust emission at \SI{0.6}{\mega\yr}. (b) Intrinsic temperature averaged along the line of sight, weighted by dust mass. (c) Standard deviation of the intrinsic temperature normalised by the average. (d) Ratio of (a) and (b).}
    \label{fig:scuba2tempslate}
\end{figure*}
These synthetic observations were calculated using dust temperatures which are decoupled from gas, apart from the collisional heat exchange term in \cref{eq:gasgraincool}. 

Dust surface brightness depends on the Planck function $B\subnu(T_d)$ and in the optically thin limit is also proportional to the dust opacity. Taking the ratio of brightnesses at two wavelengths allows the dust temperature to be calculated if the opacity spectral index is known:
\begin{equation}
	\label{eq:dustratio}
	 \frac{S_{\lambda_1}}{S_{\lambda_2}} =  \left(\frac{\lambda_2}{\lambda_1}\right)^{3+\beta} \frac{\exp(hc/\lambda_2 k T_d) - 1}{\exp(hc/\lambda_1 k T_d) - 1}
\end{equation}
where $S_{\lambda_1},S_{\lambda_2}$ are the surface brightnesses at wavelengths $\lambda_1,\lambda_2$, and $\beta$ is the dust opacity spectral index. $\beta$ is normally taken to be 2 for regions such as the one modelled here \citep[e.g.][]{reid2005,sadavoy2012,rumble2015,figueira2017} and this is the value we use.

We apply \cref{eq:dustratio} pixel-by-pixel to synthetic observations at 450 and \SI{850}{\micron}, corresponding to the Submillimetre Common User Bolometer Array (SCUBA-2) on the James Clerk Maxwell Telescope (JCMT). The resulting temperature map is presented in \cref{fig:scuba2tempsearly} and \cref{fig:scuba2tempslate}, next to an intrinsic value $\langle T_d^\textrm{int} \rangle$ which is an dust-mass-weighted average along the line of sight calculated directly from the RHD model. 

The observed temperature $\langle T_d^\textrm{obs} \rangle$ is accurately recovered in dense filaments. In warm diffuse gas, the dust temperature is underestimated by a factor of 2 and this gets worse as the \HII{} region cavity grows and becomes less dense. This is where the most massive star is located and is where discrepancies are highest. 
\citet{koepferl2017a} produced synthetic dust emission from SPH models of \citet{dale2014} and concluded that errors in temperature were present in regions which had both low density and high temperature dispersion. We agree with this at early times when the \HII{} is still mostly confined (\cref{fig:scuba2tempsearly}). At later times when the \HII{} region is extended and diffuse, inconsistencies are more correlated with low densities, with some high-temperature dispersion areas still being recovered fairly accurately (e.g. the filaments on the right of \cref{fig:scuba2tempslate}). They also note that their errors are found in places that are cooler than their surroundings and are greatest at the edge of the \HII{} region. We do not in general find this to be the case, as the differences in our model are concentrated in the interior at both early and late stages; furthermore, some cool filaments entrenched in warmer areas still have accurate temperatures. \citeauthor{koepferl2017a}'s method of SED fitting requires both the surface density and dust temperature to be free parameters, with overestimates in one leading to underestimates in the other. The method we employ here does not make any assumption on the density and therefore any discrepancies are discrepancies in temperature only. Using the ratio method only requires observations at two wavelengths, while the blackbody fitting method needs many wavelengths -- temperatures in dense regions with low temperature dispersion are accurately calculated using both methods, so in these areas observers may find the ratio method more useful due to its less stringent data requirements.

%%%%%%%%%%%%%%%%%%%%%%%%%%%
%%%%%% CONCLUSIONS
%%%%%%%%%%%%%%%%%%%%%%%%%%%
\section{Summary and conclusions}
\label{sec:conclusions}
We have modelled a \SI{1000}{\msol} cloud containing a \SI{34}{\msol} massive star including photoionizing radiation and radiation pressure feedback. In summary:
\begin{itemize}
    \item The cloud is dispersed within \SI{1.6}{\mega\yr} or $0.74\,\langle t_\textrm{ff} \rangle$, with all mass leaving the $(\SI{15.5}{pc})^3$ grid over this time. 
    \item Thermal pressure from photoionization is an efficient feedback mechanism, causing mass fluxes of the order \SI{e-3}{\msol\per\yr} at the simulation volume boundary. 
    \item At most 40 per cent of the mass gets ionized (\SI{440}{\msol}), while almost 90 per cent of the volume gets ionized. This arises from the densest filamentary structures resisting ionization, getting shovelled by the expanding ionization front into globules and pillars, which remain neutral and shield downwind material from the stellar radiation field. \item Radiation pressure plays a negligible role, causing only a slight delay in the breakout of the ionization. It is expected to be more significant at higher number densities, $n>\SI{100}{\per\cm\cubed}$ \citep{draine2011}.
    \item We use a detailed radiative transfer scheme in our models, calculating photoionization balance and thermal balance before each hydrodynamics timestep. Ionization fractions are calculated for multiple atomic species, and temperatures are calculated for gas and dust separately. These are then used to create self-consistent synthetic observations. 
    \item We also calculate the FUV interstellar radiation field, $G_0$, throughout the simulation volume, including around sink particles which can suddenly be  exposed to the flux from the massive star as gas is dispersed. 
\end{itemize}

Our synthetic observations include line and continuum emission. We have tested the use of radio free-free emission in probing the production rate of Lyman continuum photons. The `observed' rate is almost always underestimated -- by a factor of a few at early times (before \SI{0.4}{\mega\yr}), and by up to three orders of magnitude at late times (once significant amounts of mass have left the simulation volume, and thus photons would excite this gas beyond the boundary). We emphasise that radio measurements serve as a lower boundary on the production rate, especially when only part of the \HII{} region is observed and if ionizing photon escape fractions are not also measured.

We also investigated the use of brightness ratios of synthetic dust continuum at two wavelengths (450 and \SI{850}{\micron}) to probe the dust temperature. This accurately recovers the actual temperature in regions of high density and low temperature dispersion. However, in low densities or high temperature dispersions, the `observed' temperature is underestimated by a factor of 2 or more, getting worse at late stages in the very diffuse \HII{} region ($n \lesssim \SI{3}{\per\cm\cubed}$). At high densities the ratio method is as accurate as SED-fitting which requires more than two wavelengths and has surface density as an additional free parameter that can cause further discrepancies in temperature \citep{koepferl2017a}.

The cloud we have modelled here represents a cloud core as opposed to a GMC in its own right, therefore it would be embedded inside a larger-mass object, which itself would be inside an even larger-mass one. In future models we plan to go up the hierarchy to full GMC-scale regions of \SI{e6}{\msol} and \SI{100}{pc}. Currently we place stars as sink particles in dense regions, but the particles do not accrete material (though they do move around with gas and N-body interactions). For the higher-mass cloud models, we will implement a subgrid model for sink particles such that they represent clusters of stars, and will enable accretion so that they grow self-consistently.

\section*{Acknowledgements}
We thank the referee, Alejandro Raga, for helpful comments. We also thank Thomas Haworth and David Acreman for useful discussions. AA is funded by an STFC studentship. TJH and TAD are funded by STFC Consolidated Grant ST/M00127X/1. The calculations for this paper were performed on the DiRAC Complexity system at the University of Leicester and the DiRAC Data Centric system at Durham University. These form part of the STFC DiRAC HPC Facility (www.dirac.ac.uk). Complexity is funded by BIS National E-Infrastructure capital grant ST/K000373/1 and STFC DiRAC Operations grants ST/K0003259/1 and ST/M006948/1. Data Centric is funded by a BIS National E-infrastructure capital grant ST/K00042X/1, STFC capital grants ST/K00087X/1 and ST/P002307/1, DiRAC Operations grant ST/K003267/1, and Durham University. We also used the University of Exeter Supercomputer, Zen, a DiRAC Facility jointly funded by STFC, the Large Facilities Capital fund of BIS, and the University of Exeter.

%%%%%%%%%%%%%%%%%%%%%%%%%%%%%%%%%%%%%%%%%%%%%%%%%%



\bibliographystyle{mnras}
\bibliography{refs}

%%%%%%%%%%%%%%%%%%%%%%%%%%%%%%%%%%%%%%%%%%%%%%%%%%

%%%%%%%%%%%%%%%%% APPENDICES %%%%%%%%%%%%%%%%%%%%%

%\appendix


%%%%%%%%%%%%%%%%%%%%%%%%%%%%%%%%%%%%%%%%%%%%%%%%%%

% Don't change these lines
\bsp	% typesetting comment
\label{lastpage}
%\end{NoHyper} 
\end{document}